\documentclass[iicol,sn-mathphys]{sn-jnl}

\usepackage{graphicx}%
\usepackage{multirow}%
\usepackage{amsmath,amssymb,amsfonts}%
\usepackage{amsthm}%
\usepackage{mathrsfs}%
\usepackage[title]{appendix}%
\usepackage{xcolor}%
\usepackage{textcomp}%
\usepackage{manyfoot}%
\usepackage{booktabs}%
\usepackage{algorithm}%
\usepackage{algorithmicx}%
\usepackage{algpseudocode}%
\usepackage{listings}%
\usepackage{epsfig}
\usepackage{epstopdf}
\usepackage{slashed}
\usepackage{multirow,color}
\usepackage{float}
\usepackage{diagbox}
\usepackage{CJK}
\usepackage{color}
\usepackage{times}
\usepackage{subfigure}
\usepackage{bm}
\usepackage{braket}
\usepackage{array}
\usepackage[mathscr]{euscript}
\usepackage{caption}
\usepackage{makecell}
\usepackage{hyperref}
\usepackage{slashed}
\usepackage[numbers,sort&compress]{natbib}
\UseRawInputEncoding

\theoremstyle{thmstyleone}%
\theoremstyle{thmstyletwo}%

\theoremstyle{thmstylethree}%

\begin{document}

\title[Article Title]{Prospects for detecting the couplings of axion-like particle with neutrinos at the CEPC}


\author[1,2]{\fnm{Chong-Xing} \sur{Yue}}\email{cxyue@lnnu.edu.cn}

\author*[1,2]{\fnm{Xin-Yang} \sur{Li}}\email{lxy91108@163.com}

\author[1,2]{\fnm{Xiao-Chen} \sur{Sun}}\email{xcsun0315@163.com}

\affil[1]{\orgdiv{Department of Physics}, \orgname{Liaoning Normal University}, \city{Dalian}, \postcode{116029}, \country{China}}

\affil[2]{\orgdiv{Center for Theoretical and Experimental High Energy Physics}, \orgname{Liaoning Normal University}, \orgaddress{\country{China}}}


\abstract{We explore the possibility of detecting the couplings of axion-like particle~(ALP) with leptons from their loop-level impact on the ALP couplings to electroweak~(EW) gauge bosons via the signal process $e^+ e^- \to \gamma \gamma \slashed E$ at the Circular Electron Positron Collider~(CEPC) and obtain prospective sensitivities to the ALP-lepton couplings. Our numerical results show that the CEPC with the center of mass energy $\sqrt{s}=240~(91)$ GeV and the integrated luminosity $\mathcal{L}=5.6~(16)$ ab$^{-1}$ might be sensitive for probing the ALP-lepton couplings in the ALP mass range from $10$ GeV to $130~(50)$ GeV, where the prospective sensitivities to the ALP-neutrino coupling $Tr(g_{a\nu\nu})/f_a$ can be as low as $0.016~(0.012)$ GeV$^{-1}$ and $0.0031~(2.8 \times 10^{-4})$ GeV$^{-1}$ for $c_{R}=0$ and $c_{R}=c_{L}$, and to the ALP-charged lepton coupling $Tr(g_{a\ell\ell})/f_a$ can be as low as $0.035~(0.07)$ GeV$^{-1}$ for $c_{L}=0$. We find that the prospective sensitivities given by the CEPC are not covered by the current and expected exclusion regions in some ALP mass intervals. The CEPC has the potential for exploring the ALP-neutrino couplings via some promising processes induced by the ALP-EW gauge boson couplings in the future.}

\maketitle

\section{Introduction}\label{sec1}

Many new physics scenarios beyond the Standard Model~(SM) predict the existence of Axion-like particles~(ALPs), which are generalizations of the QCD axions proposed to solve the strong CP problem~\cite{Peccei:1977hh,Peccei:1977ur,Weinberg:1977ma} and arise as the massive Pseudo Nambu-Goldstone Bosons~(PNGBs) associated with spontaneously broken global symmetries at high scales. They are CP-odd scalars and gauge-singlets under the SM. The interactions of ALPs with the SM particles can be studied within the framework of Effective Field Theory~(EFT)~\cite{Georgi:1986df,Brivio:2017ije}, where the leading order interactions of ALPs with the SM particles appear at the level of dimension-five operators, suppressed by the scale $f$ at which the spontaneous breaking occurs, the shift symmetry associated with ALPs forces their couplings with fermions to be derivative and with gauge bosons involving dual field strengths. The masses of ALP and their couplings to the SM particles are considered to be independent parameters. This property makes ALPs have a much wider parameter space and  hence generate rich physical phenomenology in low- and high-energy experiments.

The masses and the couplings of ALP  can extend over many orders of magnitude, which are constrained by astrophysical and cosmological observations, as well as collider experiments. The severity of the constraints depends on the ALP mass range considered. Generally speaking, the most stringent limits on ALP couplings arise from cosmological and astrophysical bounds for very light ALPs with mass below the MeV, such as Big Bang Nucleosynthesis~(BBN), Cosmic Microwave Background~(CMB) and Supernova 1987a~\cite{Raffelt:2006cw,Marsh:2015xka}. For ALPs with MeV to hundreds of GeV scale masses, collider experiments become relevant and the constraints are alleviated. For example, the couplings of light ALP with top quarks have already been studied at the LHC via the process $p p \to t \bar{t} a $, in which the ALP is taken as the missing transverse energy~\cite{Brivio:2017ije}. Some other works on the ALP-top couplings have been studied, for recent reviews see, e.g. Refs.~\cite{Phan:2023dqw,Blasi:2023hvb,Anuar:2024myn}. The parameter space of ALP has been constrained via the process $\gamma\gamma \to a\to \gamma\gamma$ in Pb Pb collisions at the LHC~\cite{dEnterria:2021ljz}. The mass of ALP and its coupling to two photons have already been constrained using LEP data via the process $e^{+}e^{-}\rightarrow \gamma^{*} \rightarrow a\gamma\rightarrow3\gamma$~\cite{Mimasu:2014nea}. Recently, the interactions of ALP with electroweak gauge bosons have been explored at the LHC~\cite{Biswas:2023ksj,Cheung:2024qge}. Furthermore, future $e^+ e^- $ colliders like the CEPC~\cite{CEPC-SPPCStudyGroup:2015csa}, FCC-ee~\cite{FCC:2018evy}, ILC~\cite{ILC:2007bjz} and CLIC~\cite{CLICPhysicsWorkingGroup:2004qvu,Roloff:2018dqu} might give the projected bounds on the couplings of ALP with quarks, charged leptons and gauge bosons~\cite{Yue:2023mjm,Yue:2022ash,Calibbi:2022izs,Lu:2022zbe,Zhang:2021sio,Yue:2021iiu,CLIC:2018fvx}. Up to now, most phenomenological ALP analysis have concentrated on their couplings to quarks, charged leptons and gauge bosons. Nevertheless, there are relatively few studies about the couplings of ALP with neutrinos at collider experiments.

Recently, Ref.~\cite{Bonilla:2023dtf} has explored the effective ALP-neutrino and ALP-charged lepton interactions. In their work, it's assumed that ALP only couples to the SM leptons. The effective couplings of ALP with electroweak~(EW) gauge bosons $\gamma\gamma$, $W^+ W^-$, $ZZ$ and $Z\gamma$ are induced by one-loop ALP-lepton couplings. Thus, the current experimental constraints or the prospective constraints of future experiments on the ALP couplings with EW gauge bosons can be converted into the limits or prospective limits on the effective ALP-neutrino and ALP-charged lepton couplings. They have obtained constraints on the effective ALP-neutrino and ALP-charged lepton couplings through astrophysical and cosmological observations, beam dump experiments, rare meson decays and current collider experiments. In this work, we mainly investigate whether the CEPC is sensitive to the effective ALP-neutrino and ALP-charged lepton couplings. To do this, we explore the possibility of detecting the couplings of ALP with leptons from their loop-level impact on the ALP couplings to EW gauge bosons via the signal process $e^+ e^- \to \gamma \gamma \slashed E $ at the CEPC with the center of mass energy $\sqrt{s}=240$ GeV and the integrated luminosity $\mathcal{L}=$ $5.6$ ab$^{-1}$. Additionally, in order to know how much parameter space could be covered at a Z pole run before the CEPC runing at $240$ GeV, we also  analyse the signal of the process $e^+ e^- \to \gamma \gamma \slashed E $ at the CEPC with $\sqrt{s}=91$ GeV and $\mathcal{L}=$ $16$ ab$^{-1}$. Comparing our prospective sensitivities to the ALP-lepton couplings with other experimental bounds given in Ref.~\cite{Bonilla:2023dtf} and future experimental prospective sensitivities given by other literatures, our numerical results show that the prospective sensitivities given by the CEPC are not covered  by the exclusion regions given by current and future experiments within a certain mass range.

The organization of this work is as follows. In section 2, we summarize the effective Lagrangian for the interactions of ALP with the SM particles and generalise the promising production channels for the signal process $e^+ e^- \to \gamma \gamma \slashed E$ in three specific cases. In section 3, we provide a detailed analysis for the possibility of detecting the couplings of ALP with leptons based on the CEPC detector simulation and further compare our numerical results with other experimental bounds and the prospective sensitivities of future experiments on the ALP-lepton couplings, respectively. Finally conclusions are given in section 4.

\section{The theory framework}\label{sec2}

The interactions of ALP with the SM particles can be described by the effective Lagrangian that includes operators with dimension up to five~\cite{Bonilla:2021ufe,Bonilla:2022qgm,Georgi:1986df,Choi:1986zw}
\begin{eqnarray}
\begin{split}
\label{eq:2.1}
\mathcal{L}_{\text{ALP}}
	= &\frac{1}{4} \partial_\mu a \partial^\mu a - \frac{1}{2} m_a^2 a^2 +  \mathcal{L}_a^{gauge} \\ &+ \mathcal{L}_{\partial a}^{fermion}.
\end{split}
\end{eqnarray}
Where $\mathcal{L}_a^{gauge}$ stands for the anomalous ALP-gauge boson effective couplings, which at energies above EW symmetry breaking scale can be written as
\begin{eqnarray}
\begin{split}
\label{eq:2.2}
\mathcal{L}_a^{gauge}
	= &- C_{\tilde{G}} \frac{a}{f_a} G^{\alpha}_{\mu\nu} \tilde{G}^{\mu\nu,\alpha} - C_{\tilde{W}} \frac{a}{f_a} W^{i}_{\mu\nu} \tilde{W}^{\mu\nu,i} \\
&- C_{\tilde{B}} \frac{a}{f_a} B_{\mu\nu} \tilde{B}^{\mu\nu}.
\end{split}
\end{eqnarray}
Here $G^{\alpha}_{\mu\nu}$, $W^{i}_{\mu\nu}$ and $B_{\mu\nu}$ are respectively the $SU(3)_{C}$, $SU(2)_{L}$ and $U(1)_{Y}$ gauge field strengths, while $\tilde{G}^{\mu\nu,\alpha}$, $\tilde{W}^{\mu\nu,i}$ and $\tilde{B}^{\mu\nu}$ are the corresponding dual field strengths which are defined as $\tilde{V}^{\mu\nu}=\frac{1}{2}\epsilon^{\mu\nu\lambda\kappa}V_{\lambda\kappa}$ $(V \in G$, $W$, $B)$ with $\epsilon^{0123}= 1$. $\mathcal{L}_{\partial a}^{fermion}$ stands for the gauge-invariant ALP-fermion derivative interactions,
\begin{eqnarray}
\begin{split}
\label{eq:2.3}
\mathcal{L}_{\partial a}^{fermion}
	= \sum_{\substack{\psi}} \frac{\partial_{\mu} a}{f_a}  \bar\psi \gamma^\mu \mathbf{c}_\psi \psi,
\end{split}
\end{eqnarray}
where $\mathbf{c}_\Psi$ are $3\times3$ Hermitian matrices in flavour space. In this paper, we focus on purely leptonic couplings, then the gauge-invariant ALP-fermion derivative interactions can be written as~\cite{Bonilla:2023dtf}

\begin{equation} \label{eq:2.4}
\begin{aligned}
      \mathcal{L}&_{\partial a}^{fermion}  = \frac{\partial_\mu a}{f_a} \, \bar{L}_L \gamma^\mu c_L L_L + \frac{\partial_\mu a}{f_a} \, \bar{e}_R \gamma^\mu c_R e_R \,, \\
    & = \frac{\partial_\mu a}{f_a} \, \bar{\nu}_L \gamma^\mu c_L \nu_L + \frac{\partial_\mu a}{f_a} \, \bar{e} \gamma^\mu (c_R P_R + c_L P_L ) e \,, \\
    & = \frac{\partial_\mu a}{2 f_a} \, \bar{\nu}_L \gamma^\mu (1 - \gamma^5) c_L \nu_{\ell} + \frac{\partial_\mu a}{2 f_a} \, \bar{e} \gamma^\mu (( c_R + c_L ) \\& + ( c_R - c_L ) \gamma^5 ) e \,.
\end{aligned}
\end{equation}
Here $P_{R,L}$ stand for the chirality projectors, the left-handed EW lepton fields have been decomposed on their charged and neutral components $L_L=(e_L, \nu_L)$, with $e= e_R + e_L$. As shown in Eq.~(\ref{eq:2.4}), the ALP-neutrino couplings are defined only in terms of the coupling parameter $c_L$.

In the flavour-diagonal scenario, the "phenomenological" couplings to the physical leptons only depend on the axial component of the operators in Eq.~\eqref{eq:2.4} at tree-level~\cite{Bonilla:2021ufe,Bonilla:2022qgm}, then the ALP-fermion effective interactions can be written as
\begin{equation}\label{eq:2.5}
\begin{aligned}
    \mathcal{L}_{\partial a}^{fermion}  &\supset  \frac{\partial_\mu a}{f_a} \sum_{\ell} (g_{a\nu\nu})_{\ell\ell} \overline{\nu}_\ell \gamma^\mu \gamma^5 \nu_\ell \\
    & + \frac{\partial_\mu a}{f_a} \sum_{\ell} (g_{a\ell\ell})_{\ell\ell} \overline{\ell} \gamma^\mu \gamma^5 \ell \,,
 \end{aligned}
\end{equation}
where $g_{a\ell\ell}$ and $g_{a\nu\nu}$ are defined as
\begin{equation} \label{eq:2.6}
    g_{a\nu\nu} \equiv (- c_L) \,, \qquad g_{a\ell\ell} \equiv (c_R - c_L) \,.
\end{equation}
Indeed, the vectorial coupling vanishes due to the conservation of vectorial currents at classical level. However, at the quantum level, the lepton number currents are broken due to chiral anomalies, the vector components do contribute even in the flavour-diagonal case and thus they will induce interactions between ALP and heavy gauge bosons~\cite{Bonilla:2023dtf}.

The low-energy equations of motion (EOM) for the lepton fields including one-loop anomalous corrections can be written as ~\cite{Bonilla:2023dtf}
\begin{equation}
    \begin{aligned}
    \frac{\partial_\mu a}{f_a}\Bar{e}_R \gamma^\mu c_R& e_R =-( i\frac{a}{f_a} \Bar{e}_L M_R c_R e_R  + \text{h.c.} ) + \\& \text{Tr}\left[c_R \right]  \frac{a}{f_a} \frac{g'^2}{16 \pi^2} B_{\mu\nu}\Tilde{B}^{\mu\nu}\,, \\
    \frac{\partial_\mu a}{f_a}\Bar{e}_L \gamma^\mu c_L &e_L = ( i\frac{a}{f_a} \Bar{e}_L M_R c_L e_R  + \text{h.c.} ) -  \text{Tr}\left[c_L \right]  \\& \frac{a}{f_a} \Big[ \frac{g'^2}{64 \pi^2} B_{\mu\nu}\Tilde{B}^{\mu\nu} + \frac{g^2}{64 \pi^2} W_{\mu\nu}\Tilde{W}^{\mu\nu} \Big] \,, \\
    \frac{\partial_\mu a}{f_a}\Bar{\nu}_L \gamma^\mu c_L& \nu_L  = ( i\frac{a}{f_a} \Bar{\nu}_L M_\nu c_L \nu_R  + \text{h.c.} ) - \text{Tr}\left[c_L \right] \\ & \frac{a}{f_a}  \Big[ \frac{g'^2}{64 \pi^2} B_{\mu\nu}\Tilde{B}^{\mu\nu} + \frac{g^2}{64 \pi^2} W_{\mu\nu}\Tilde{W}^{\mu\nu} \Big] \,, \label{eq:2.7}
    \end{aligned}
    \end{equation}
with
\begin{equation} \label{eq:2.8}
    \begin{aligned}
       g=\frac{e}{s_w} \,, \qquad \qquad g'=\frac{e}{c_w} \,.
    \end{aligned}
\end{equation}
Where $g$ and $g'$ represent the $SU(2)_L$ and $U(1)_Y$ coupling constants, $M_R$ and $M_{\nu}$ denote the generic mass matrices for charged leptons and neutrinos, respectively. $c_w (s_w)$ is the cosine (sine) of the weak mixing angle $\theta_{w}$.

The above equation is comprised of two terms. The first term represents the couplings of ALP with leptons at tree-level and the second term represents the one-loop anomalous corrections from  EW gauge bosons. As shown in Eq.~(\ref{eq:2.7}), one can see that experimental bounds on ALP couplings to massive EW gauge bosons can translate into limits on the flavour-diagonal components of ALP-neutrino couplings.

After electroweak symmetry breaking, the anomalous ALP-gauge boson effective couplings given by Eq.~(\ref{eq:2.2}) can be rewritten in terms of the EW gauge boson mass eigenstates,
\begin{equation} \label{eq:2.9}
\begin{aligned}
 \mathcal{L}_a^{gauge}\,= & -\frac{1}{4} g_{agg} \,a\, G_{\mu \nu}^\alpha \widetilde{G}^{\mu \nu,\, \alpha}  - \frac{1}{4} g_{a\gamma\gamma} \,a\,F_{\mu \nu} \widetilde{F}^{\mu \nu} \\
    & - \frac{1}{4} g_{a\gamma Z} \,a\,F_{\mu \nu} \widetilde{Z}^{\mu \nu} - \frac{1}{4} g_{aZZ} \,a\,Z_{\mu \nu} \widetilde{Z}^{\mu \nu} \\
    & - \frac{1}{2} g_{aWW} \,a\,W_{\mu \nu}^+ \widetilde{W}^{\mu \nu,\,-} \,,
\end{aligned}
\end{equation}
with
\begin{equation} \label{eq:2.10}
    \begin{aligned}
        & g_{agg} \equiv \frac{4}{f_a} c_{\tilde{G}} \,,  \qquad g_{a\gamma\gamma}  \equiv \frac{4}{f_a} \left( c_w^2 c_{\tilde{B}} + s_w^2 c_{\tilde{W}} \right) \,,   & \\
        & g_{aWW}   \equiv \frac{4}{f_a} c_{\tilde{W}} \,,  \qquad  g_{aZZ}  \equiv \frac{4}{f_a} \left( s_w^2 c_{\tilde{B}} + c_w^2 c_{\tilde{W}} \right) \,,  \qquad \\
        & g_{a\gamma Z} \equiv \frac{8}{f_a} c_w s_w \left( c_{\tilde{W}} - c_{\tilde{B}} \right) \,.
    \end{aligned}
\end{equation}

\begin{figure}[!ht]
\centerline{
\includegraphics[width=0.4\textwidth]{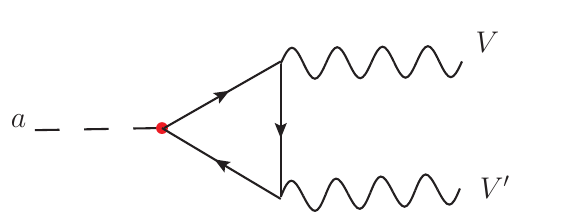}}
\vspace*{8pt}
\caption{One-loop diagram contributing to the ALP-EW gauge boson effective couplings originated from the ALP-lepton couplings.
\protect\label{fig:1}}
\end{figure}

From above discussions, we can see that the ALP-lepton couplings can induce the ALP-EW gauge boson effective couplings at one-loop, which have been carefully studied in Ref.~\cite{Bonilla:2023dtf}. The relevant Feynman diagrams are shown in Fig.~\ref{fig:1}, where $V$ and $V'$ stand for the EW gauge bosons $\gamma$, $Z$ or $W$ and the leptons running
in the loop may be charged leptons or neutrinos depending on $V$ and $V'$. The contributions can be divided into two parts. One is the chiral anomalous term of the triangle diagram, which is mass-independent, and the other one is the mass-dependent term. If the ALP mass $m_a$ is larger than the mass $m_{\ell}$ of the lepton running in the loop, then the ALP-EW gauge boson effective couplings induced by the ALP-lepton couplings can be approximately written as~\footnote{Eq.~\eqref{eq:2.11} neglects the mass-dependent one-loop contributions to the ALP-EW gauge boson effective couplings. Since the masses of the $Z$ and $W$ boson are much larger than that of the SM leptons, the mass-dependent terms provide at most corrections of order $\mathcal{O} (m_{\ell}^2 / M_Z^2)$ for $g_{a\gamma Z}^{eff}$ and $g_{aZZ}^{eff}$, and corrections of order $\mathcal{O} (m_{\ell}^2 / M_W^2)$ for $g_{aWW}^{eff}$, which can be safely ignored. Moreover, the effective ALP-photon coupling $g_{a\gamma \gamma}^{eff}$ only depends on the ratio $m_{\ell}^2/m_a^2$, when $m_a$ $\gg$ $m_{\ell}$, the mass-dependent terms become negligible. For detailed discussion, see Ref.~\cite{Bonilla:2023dtf}.}~\cite{Bonilla:2023dtf}

\begin{equation}\label{eq:2.11}
\begin{aligned}
 &
 g_{a\gamma\gamma}^{eff} \approx - \frac{\alpha_{em}}{\pi f_a}Tr(g_{a\ell\ell}),
 \\
 &
 g_{aWW}^{eff} \approx - \frac{\alpha_{em}}{2s_w^2 \pi f_a} Tr(g_{a\nu\nu}),
 \\
 &
 g_{a\gamma Z}^{eff} \approx - \frac{\alpha_{em}}{s_w c_w \pi f_a}[2 s_w^2 Tr(g_{a\ell\ell})  - Tr(g_{a\nu\nu})],
 \\
 &
 g_{aZZ}^{eff} \approx - \frac{\alpha_{em}}{2 s_w^2 c_w^2 \pi f_a}[2 s_w^4 Tr(g_{a\ell\ell})+ \\& \qquad \qquad
 (1- 2 s_w^2)Tr(g_{a\nu\nu})],
\end{aligned}
\end{equation}
with
\begin{equation} \label{eq:2.12}
    \begin{aligned}
       \alpha_{em}=\frac{e^2}{4\pi} = \frac{g^2 g'^2}{4\pi(g^2+g'^2)} \,.
    \end{aligned}
\end{equation}

One can see from above equations that the experimental bounds on the ALP-photon effective coupling $g_{a\gamma \gamma}^{eff}$ lead to bounds on the coupling $Tr(g_{a\ell\ell})$ only, while the experimental bounds on the ALP-$WW$ effective coupling $g_{aWW}^{eff}$ can be directly translated into bounds on the coupling $Tr(g_{a\nu\nu})$. The experimental bounds on the ALP-$\gamma Z$ effective coupling $g_{a\gamma Z}^{eff}$ and ALP-$ZZ$ effective coupling $g_{aZZ}^{eff}$ lead to bounds on two different combinations of both couplings defined in Eq.~\eqref{eq:2.6}. In this work, we mainly investigate the possibility of detecting the couplings of ALP with leptons via the signal process $e^+ e^- \to \gamma \gamma \slashed E $ at the CEPC with $\sqrt{s}=240$~($91$) GeV and $\mathcal{L}= 5.6$~($16$) ab$^{-1}$. In our discussion, the mass range of ALPs from $10$ GeV to $130$~(50) GeV is considered, which is much larger than the mass of the SM leptons $m_{\ell}$. Then the signal can be specifically analyzed using Eq.~\eqref{eq:2.11}.

For a very light ALP with $m_a < 2 m_e$, $a\to \gamma\gamma$ is the only allowed decay channel, and if the ALP mass is larger than $2 m_e$ $\approx$ $1.022$ MeV, the leptonic decay $a\to \ell^+ \ell^-$ can be the dominant ALP decay channels if kinematically allowed. When the ALP mass $m_a$ $>$ $90$ GeV, the $Z\gamma$ mode will open up. The relevant decay widths of ALP as a function of the ALP mass $m_a$ and the corresponding coupling coefficient are given by Eq.~\eqref{eq:2.13}~\cite{Bauer:2017ris,Bauer:2020jbp,Craig:2018kne,Bonilla:2021ufe}.

\begin{equation}\label{eq:2.13}
\begin{aligned}
 &
 \Gamma(a\to\ell^{+}\ell^{-}) =  \frac{m_a m_{\ell}^2}{8\pi}|g_{a\ell\ell}^{eff}|^2 \sqrt{1-\frac{4m_{\ell}^2}{m_a^2}},
 \\
 &
 \Gamma(a\to\gamma\gamma) =  \frac{m_a^3}{64\pi}|g_{a\gamma\gamma}^{eff}|^2,
 \\
 &
  \Gamma(a\to Z \gamma) =  \frac{m_a^3}{128\pi}|g_{a Z \gamma}^{eff}|^2 (1-\frac{m_Z^2}{m_a^2})^3.
\end{aligned}
\end{equation}

\begin{figure*}[!ht]	
\begin{center}
\subfigure[]{\includegraphics [scale=0.28]{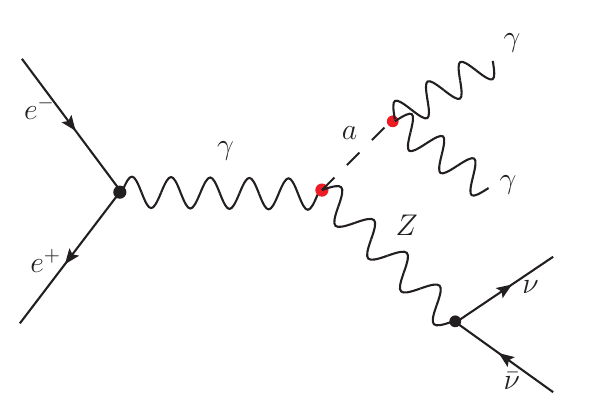}}
\hspace{0.2em}
\subfigure[]{\includegraphics [scale=0.28]{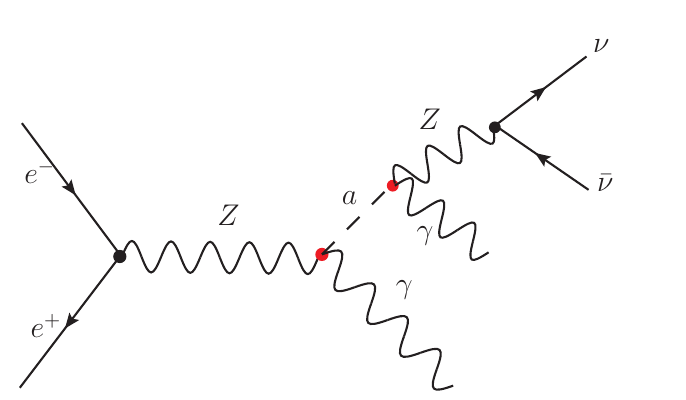}}
\hspace{0.2em}
\subfigure[]{\includegraphics [scale=0.28]{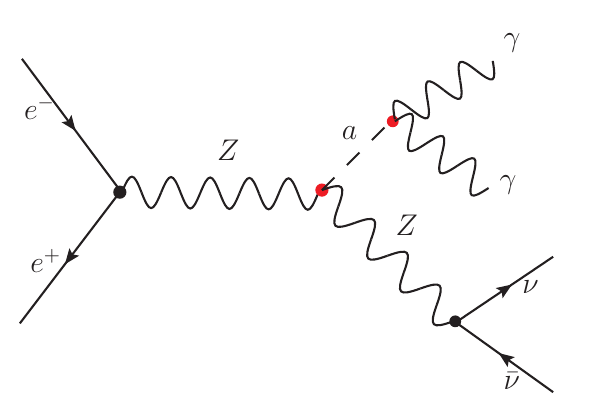}}
\subfigure[]{\includegraphics [scale=0.28]{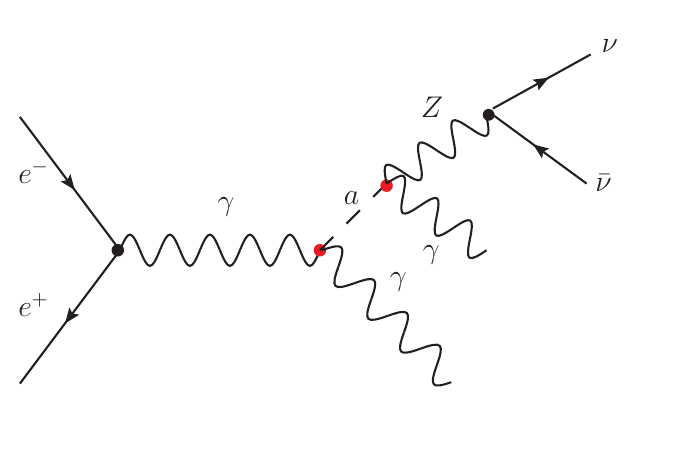}}
\subfigure[]{\includegraphics [scale=0.28]{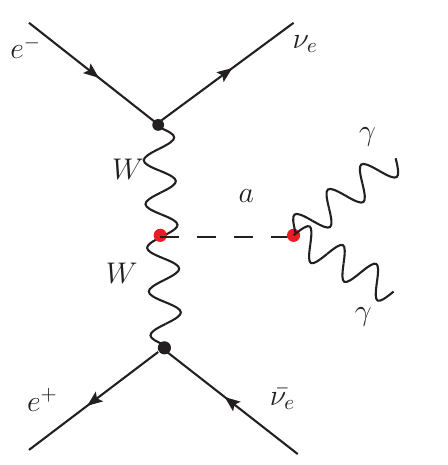}}
\caption{The Feynman diagrams for the signal process $e^+ e^- \to \gamma \gamma \slashed E $ induced by ALPs. }
\label{fig:2}
\end{center}
\end{figure*}

The Feynman diagrams for the process $e^+ e^- \to \gamma \gamma \slashed E $ induced by ALPs have been shown in Fig.~\ref{fig:2}. For $c_{R}=0$, ALP has the same coupling strength to charged leptons and neutrinos, i.e. $Tr(g_{a\ell\ell})=Tr(g_{a\nu\nu})=-c_L$, in which case the ALP-neutrino and ALP-charged lepton couplings can be explored via all the channels in Fig.~\ref{fig:2}. For $c_{L}=0$, there is no ALP-neutrino couplings, i.e. $Tr(g_{a\nu\nu})=0$, in which case $W^+ W^-$ fusion process is absent and the ALP-charged lepton couplings can be explored via the first four channels in Fig.~\ref{fig:2}. For $c_{L}=c_{R}$, there is no tree-level flavour-diagonal ALP-charged lepton interactions, i.e. $Tr(g_{a\ell\ell})=0$, in which case ALP cannot couple to two photons, which is known as photophobic ALP~\cite{Craig:2018kne}. In this case, the ALP-neutrino couplings can be explored only via the channel shown in Fig.~\ref{fig:2}{b}. In next section, we will specifically analyze these three cases, i.e. $c_{R}=0$, $c_{L}=0$ and $c_{L}=c_{R}$.

\section{The possibility of detecting the couplings of ALP with leptons at the CEPC}
\begin{figure*}[!ht]
\centering
\subfigure[]{\includegraphics [scale=0.25]{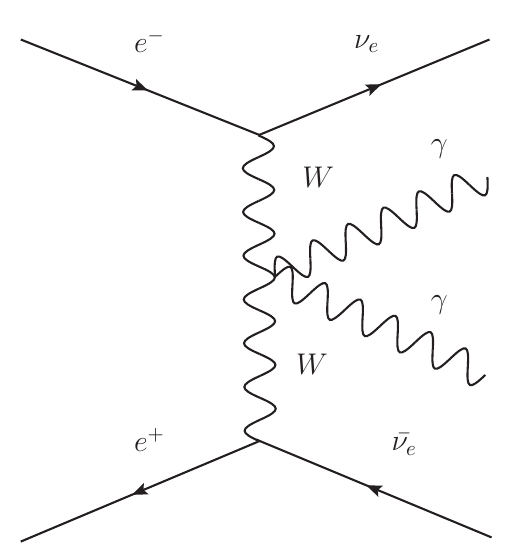}}
\hspace{0.2em}
\subfigure[]{\includegraphics [scale=0.25]{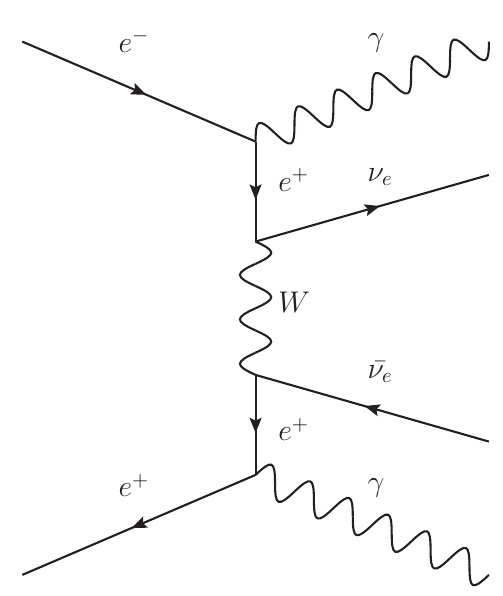}}
\hspace{0.2em}
\subfigure[]{\includegraphics [scale=0.25]{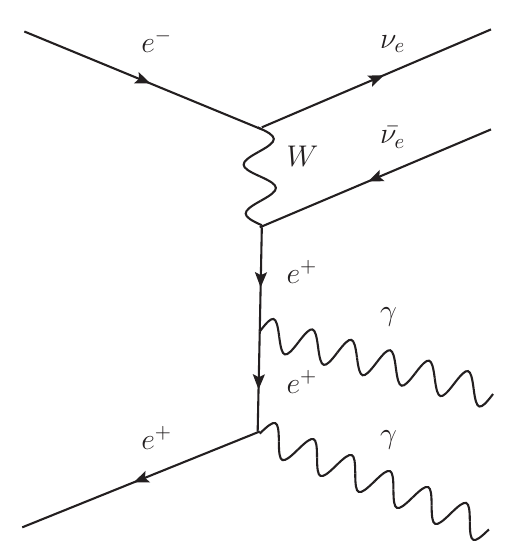}}
\subfigure[]{\includegraphics [scale=0.25]{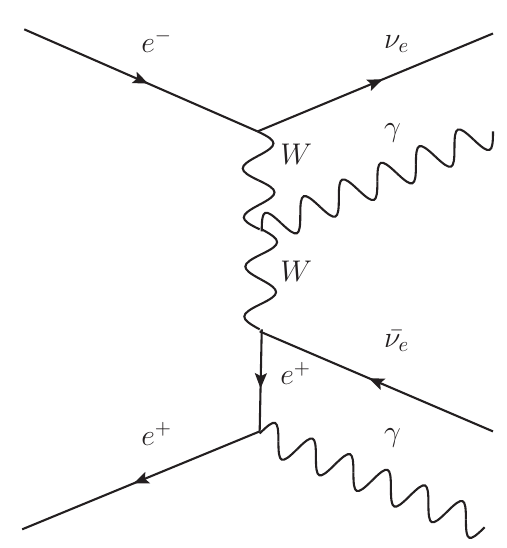}}
\subfigure[]{\includegraphics [scale=0.25]{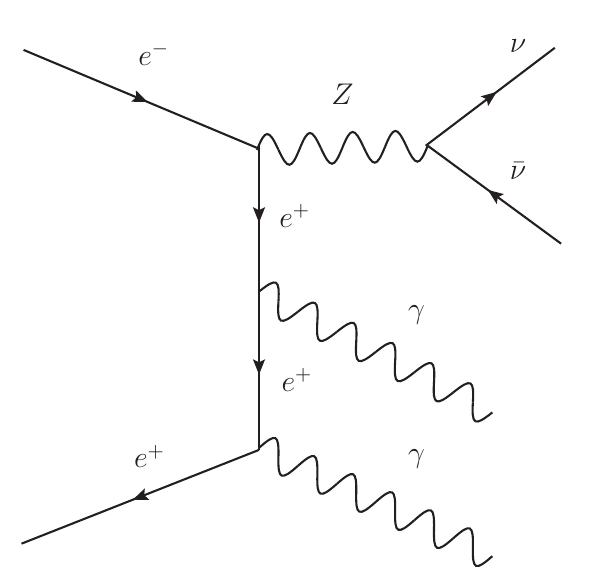}}
\subfigure[]{\includegraphics [scale=0.25]{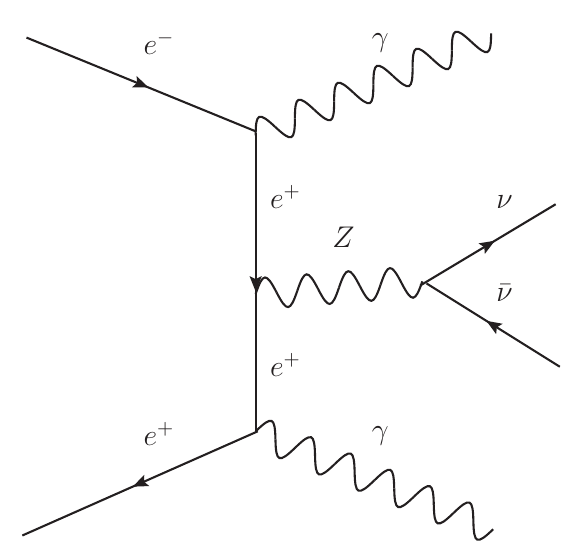}}
\caption{The typical Feynman diagrams for the SM background of the process $e^+ e^- \to \gamma \gamma \slashed E $. }
\label{fig:3}
\end{figure*}

We will explore the possibility of detecting the couplings of ALP with leptons via the process $e^+ e^- \to \gamma \gamma \slashed E $ as shown in Fig.~\ref{fig:2} at the $240$~($91$) GeV CEPC with $\mathcal{L}=$ $5.6$~$(16)$ ab$^{-1}$ in three cases mentioned above with the ALP mass $m_a$ from $10$ GeV to $130$~($50$) GeV. The main SM background considered in our numerical analysis is the process $e^+ e^- \to \gamma \gamma \slashed E$  shown in Fig.~\ref{fig:3}, which are mainly induced by electroweak interaction. The contribution from the $\nu_e \bar{\nu_e} \gamma \gamma$ final state mediated by $W$ boson, in which two photons are radiated separately from the interior of the $W$ boson, is subordinate and is not included in Fig.~\ref{fig:3}, which is also taken into consideration in our work. In addition, the process $e^+ e^- \to \gamma \gamma \nu \bar{\nu} \nu \bar{\nu}$ can also contribute to the SM background. However, they can be safely ignored as their contributions to SM background are estimated to be less than $0.004$$\%$.

We use \verb"FeynRules"~\cite{Alloul:2013bka} to generate the model file for the effective Lagrangian. A Monte Carlo~(MC) simulation is performed to explore the possibility of detecting the couplings of ALP with leptons at the CEPC. All the signal and background that are invisible neutrinos and two photons going to be discussed in the next subsections will be generated in \verb"MadGraph5_aMC@NLO"~\cite{Alwall:2014hca} with basic cuts. In other words, in our numerical calculation, we require the photons with transverse momentum $p_{T}^{\gamma}>10$ GeV. The absolute value of the photon pseudorapidity $\eta_{\gamma}$ needs to be less than $2.5$. The angular separation requirement is $\Delta R_{\gamma\gamma}>0.4$ for two photons, in which $\Delta$$R$ is defined as $\sqrt{(\Delta \phi)^{2}+(\Delta \eta)^{2}}$. The \verb"PYTHIA8" program~\cite{Sjostrand:2014zea} is implemented for showering, and we use \verb"DELPHES"~\cite{deFavereau:2013fsa} for the fast simulation of the CEPC detector and \verb"MadAnalysis5"~\cite{Conte:2012fm,Conte:2014zja,Conte:2018vmg} for the analysis of the resulting output.

\subsection{In the case $c_{R}=0$, $Tr(g_{a\ell\ell})=Tr(g_{a\nu\nu})$}\label{subsec1}

\begin{figure}[h]
\centerline{
\includegraphics[width=0.5\textwidth]{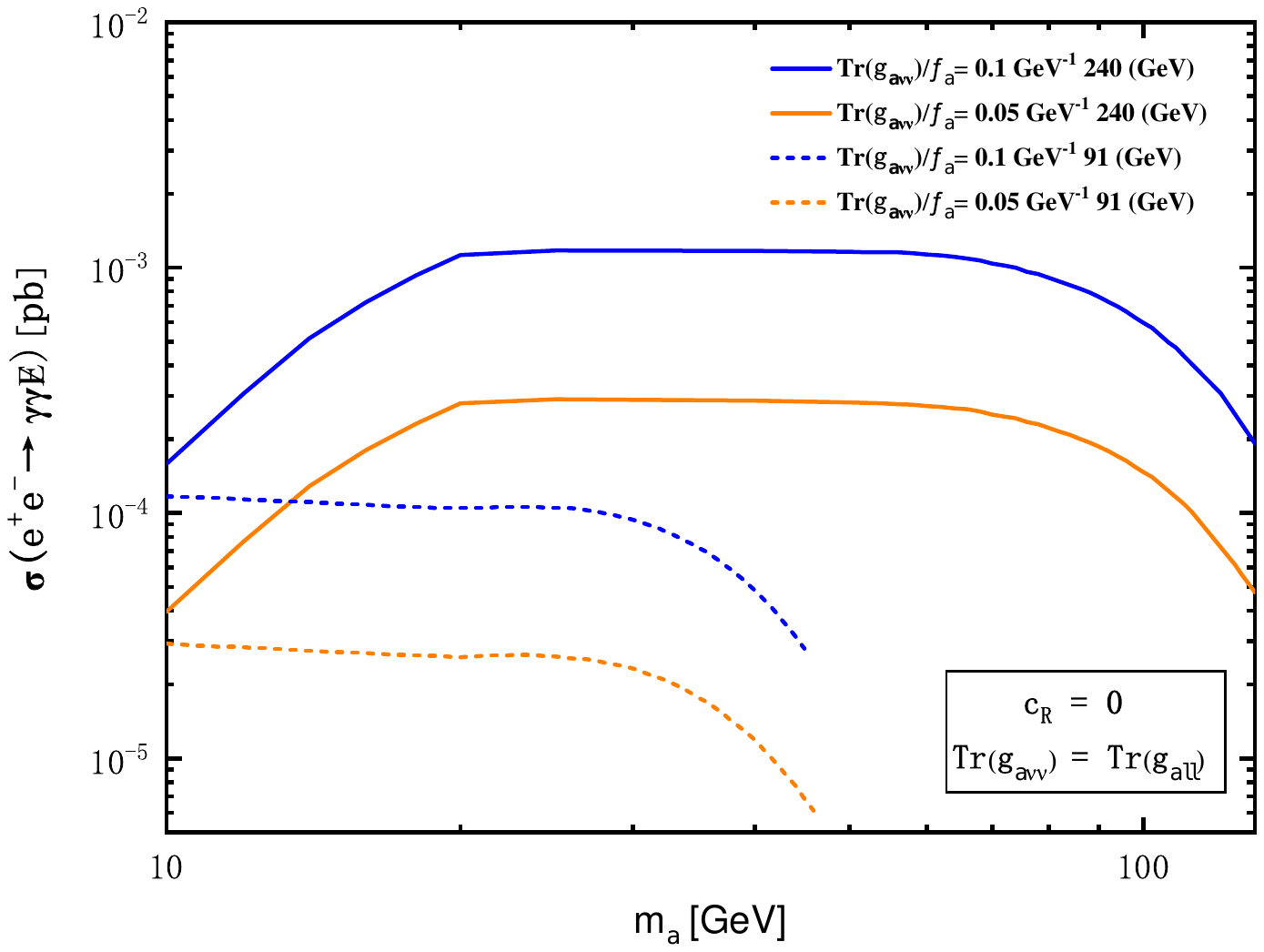}}
\vspace*{8pt}
\caption{The production cross section of the signal process $e^{+}e^{-}\rightarrow\gamma\gamma\slashed E$ as a function of the ALP mass $m_a$ at the $240$ GeV~(solid line) and the $91$ GeV~(dashed line) CEPC for $c_{R}=0$.
\protect\label{fig:4}}
\end{figure}

\begin{figure*}[!ht]
\centering
\subfigure[]{\includegraphics [scale=0.25]{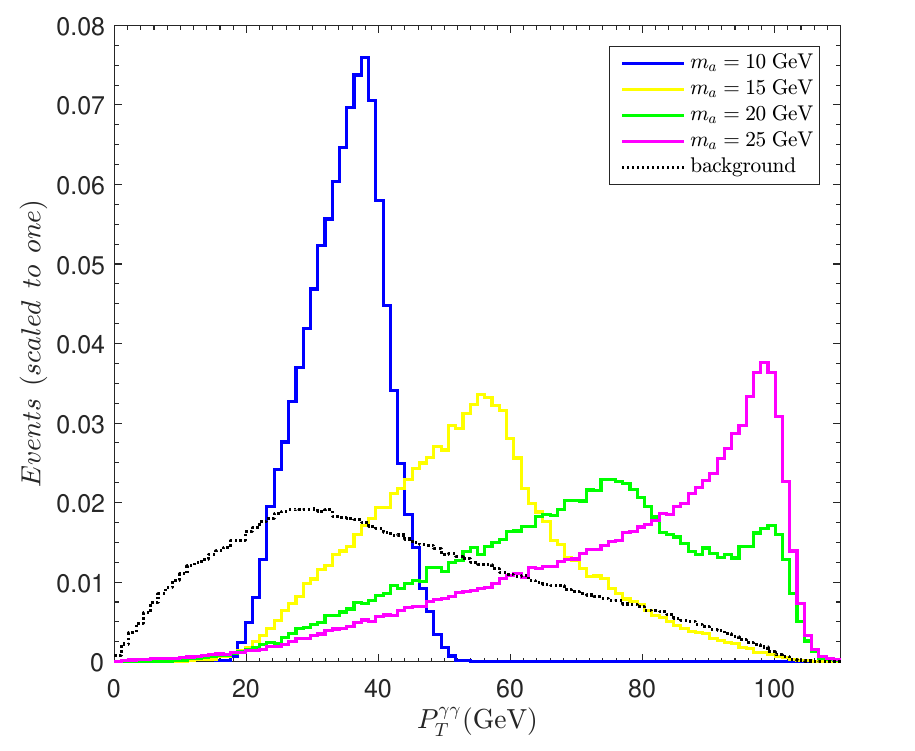}}
\hspace{0.2em}
\subfigure[]{\includegraphics [scale=0.25]{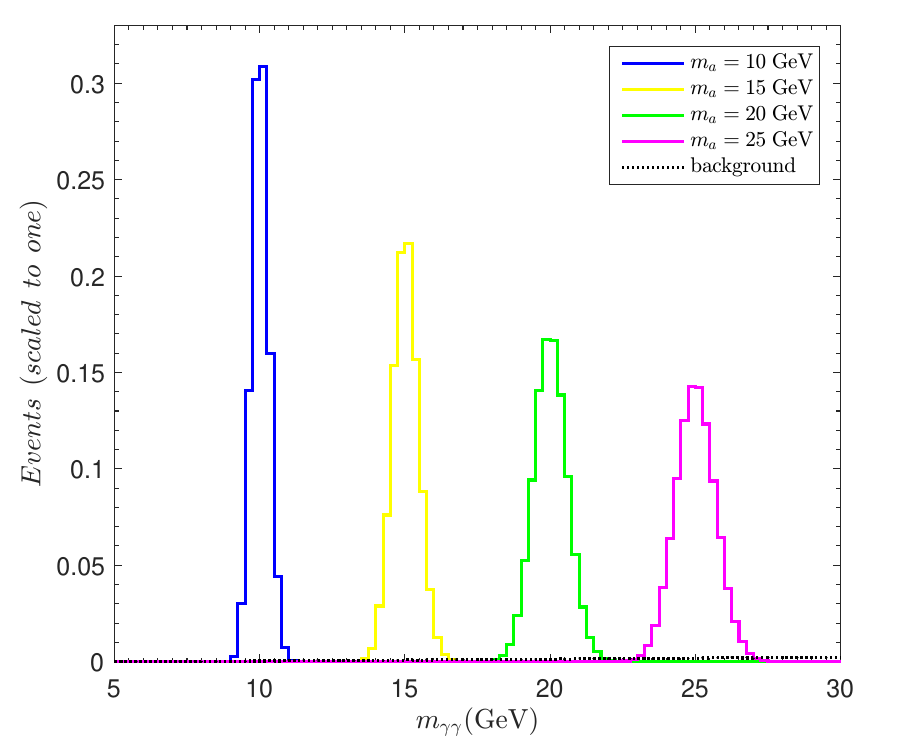}}
\hspace{0.2em}
\subfigure[]{\includegraphics [scale=0.25]{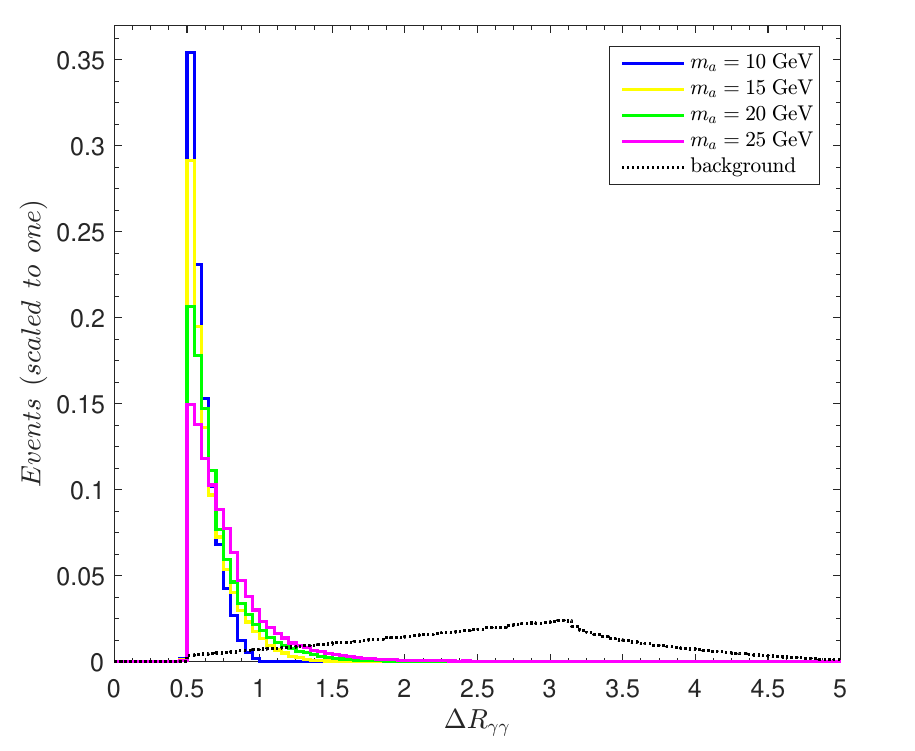}}
\caption{The normalized distributions of the observables $p_{T}^{\gamma\gamma}$ (a), $m_{\gamma\gamma}$ (b) and $\Delta R_{\gamma\gamma}$ (c) for the signal of selected ALP-mass benchmark points and SM background at the 240 GeV CEPC with $\mathcal{L}=$ $5.6$ ab$^{-1}$.}
\label{distribution:1}
\end{figure*}

\begin{figure*}[!ht]
\centering
\subfigure[]{\includegraphics [scale=0.25] {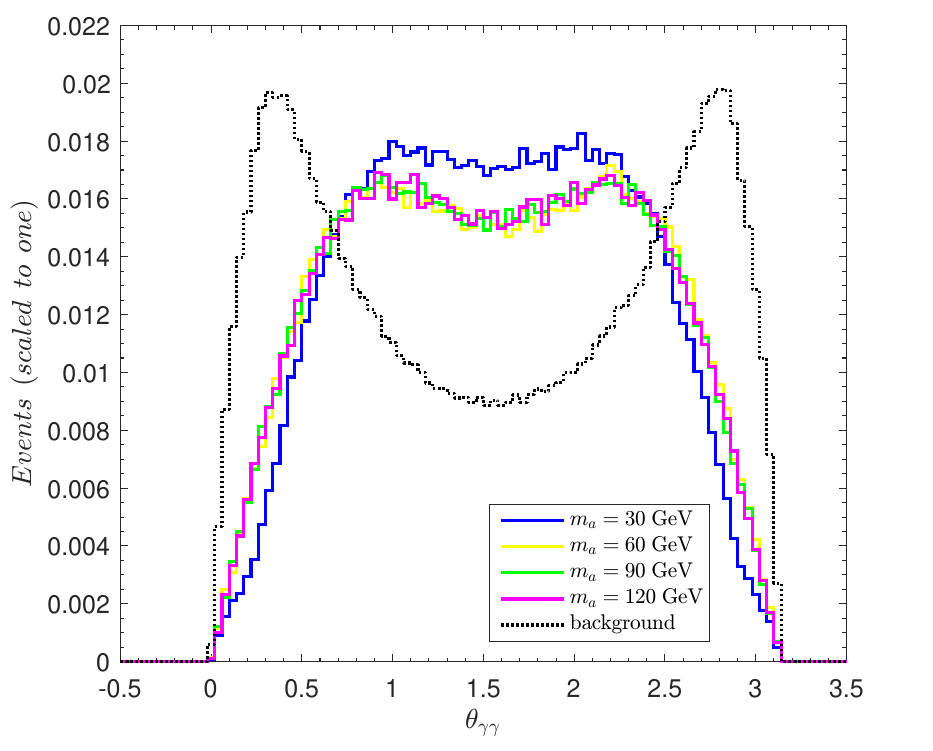}}
\subfigure[]{\includegraphics [scale=0.25] {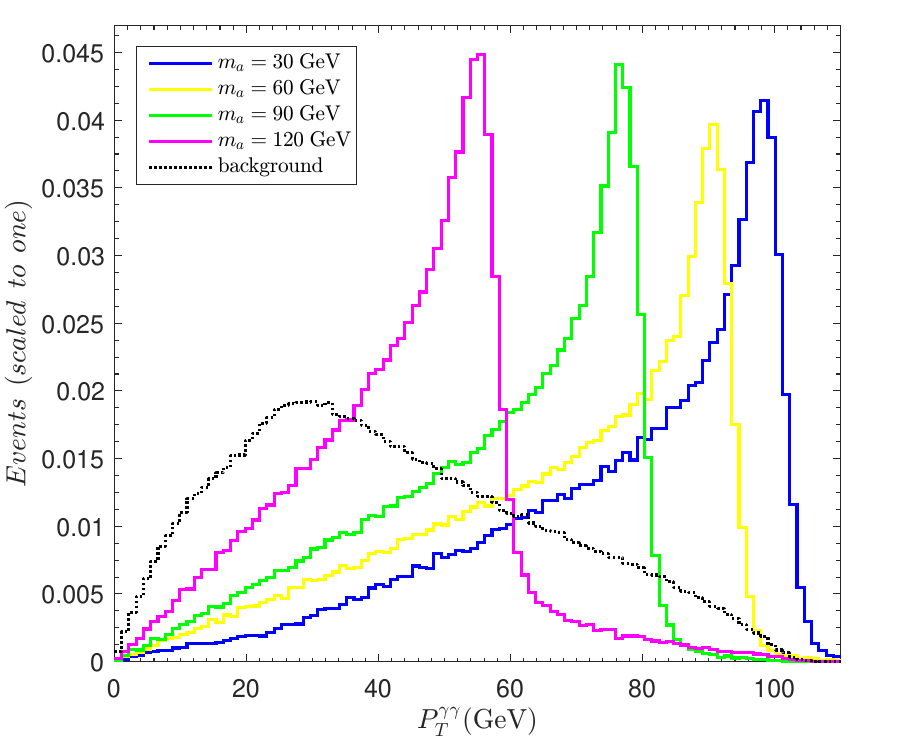}}
\subfigure[]{\includegraphics [scale=0.25] {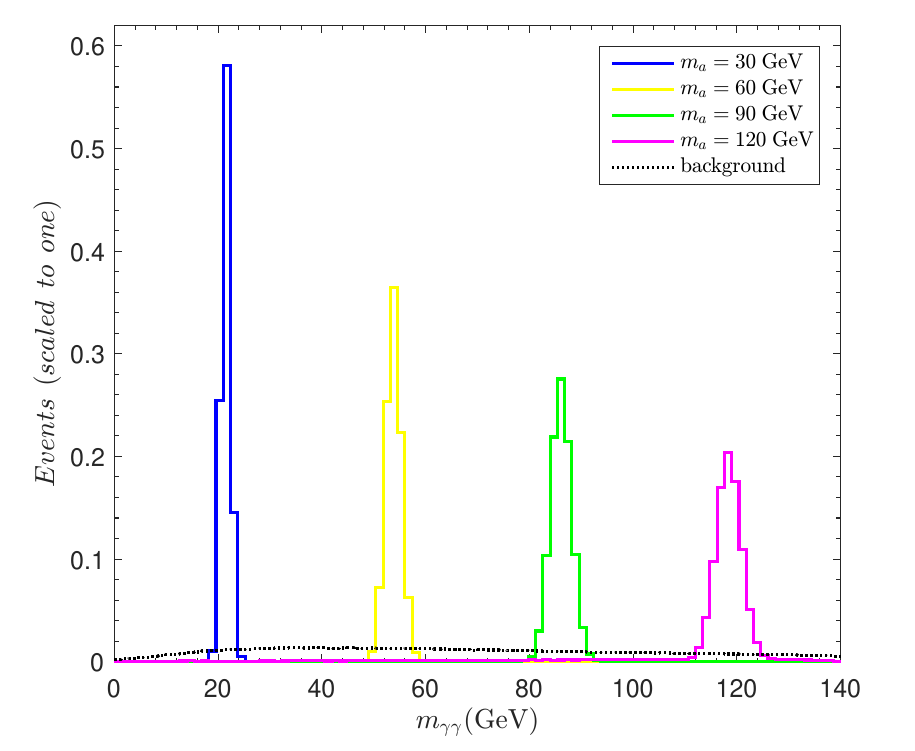}}
\subfigure[]{\includegraphics [scale=0.25] {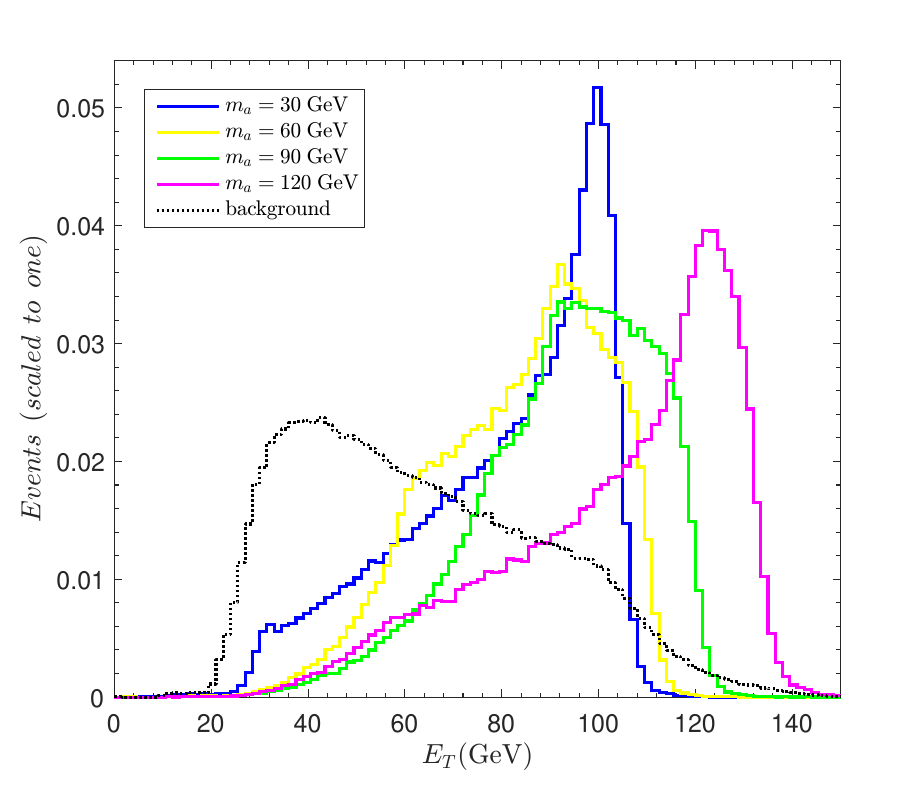}}
\caption{The normalized distributions of the observables $\theta_{\gamma\gamma}$ (a), $p_{T}^{\gamma\gamma}$ (b), $m_{\gamma\gamma}$ (c) and  $E_T$ (d) for the signal of selected ALP-mass benchmark points and SM background at the 240 GeV CEPC with $\mathcal{L}=$ $5.6$ ab$^{-1}$.}
\label{distribution:2}
\end{figure*}

Now let's begin with the process $e^+ e^- \to \gamma \gamma \slashed E $ induced by ALPs for $c_{R}=0$, $Tr(g_{a\ell\ell})=Tr(g_{a\nu\nu})$. In this case, all the diagrams in Fig.~\ref{fig:2} have contributions. The production cross section of the signal process $e^{+}e^{-}\rightarrow\gamma\gamma\slashed E$ induced by ALPs with $10$ GeV $\leq m_a \leq$ $130$~$(50)$ GeV and different $Tr(g_{a\nu\nu})/f_a$ at the $240~(91)$ GeV CEPC has been shown in Fig.~\ref{fig:4}. As illustrated in Fig.~\ref{fig:4}, the cross section of the signal process $e^{+}e^{-}\rightarrow\gamma\gamma\slashed E$ induced by ALPs at the $240$ GeV CEPC increases as the coupling coefficient $Tr(g_{a\nu\nu})/f_a$ increasing and reaches peak at the ALP mass of about $25$ GeV, which is because the final state phase space becomes limited with increasing ALP mass. At this point, the values of cross section are $2.917 \times 10^{-4}$ pb and $1.175 \times 10^{-3}$ pb for $Tr(g_{a\nu\nu})/f_a$ equaling to $0.05$ GeV$^{-1}$ and $0.1$ GeV$^{-1}$, respectively. The signal in the considered parameter region is smaller than the corresponding SM background~($0.06727$ pb). While the cross section at the $91$ GeV CEPC decreases slowly until $m_a$ is approximately equal to $25$ GeV, and decreases rapidly when $m_a$ is greater than $25$ GeV due to the depression of the final state phase space. At this point, the values of the cross section are $2.615 \times 10^{-5}$ pb and $1.055 \times 10^{-4}$ pb for $Tr(g_{a\nu\nu})/f_a$ equaling to $0.05$ GeV$^{-1}$ and $0.1$ GeV$^{-1}$, respectively. The signal in the considered parameter region is slightly less than the corresponding SM background~($1.311 \times 10^{-4}$ pb). Based on the changing trend of the production cross section at the $240$ ($91$) GeV CEPC, we take $25$ GeV as a breakpoint to divide the mass range of ALP considered into two intervals for the study. For the ALP with mass in the range of $10$ GeV to $25$ GeV, kinematic variables of the transverse momentum of reconstructed ALP $p_{T}^{\gamma\gamma}$, invariant mass distribution of two photons $m_{\gamma\gamma}$ and the angular separation between two photons $\Delta R_{\gamma\gamma}$ are chosen. We present in Fig.~\ref{distribution:1} the distributions of $p_{T}^{\gamma\gamma}$, $m_{\gamma\gamma}$ and $\Delta R_{\gamma\gamma}$ for the signal and background events with typical points of $m_a = 10$, $15$, $20$, $25$ GeV at the $240$ GeV CEPC with $\mathcal{L}= 5.6$ ab$^{-1}$.
The two photons from the light ALP decay is much closer, whereas the angular separation between them tends to have a wide distribution for the background events. Besides, it can be seen that the signal and background can be well distinguished in the invariant mass $m_{\gamma\gamma}$ distribution. The signal events have peaks around the ALP mass, which become wider as the ALP mass increases. Therefore, for the range of $m_a$ between $10$ GeV to $25$ GeV, we employ the optimized mass window cut $\lvert m_{\gamma\gamma}- m_a\rvert\leq3$ GeV, while for the range of $m_a$ between $25$ GeV to $130~(50)$ GeV, we employ the optimized mass window cut $\lvert m_{\gamma\gamma}- m_a\rvert\leq5$ GeV below. Variables of the angle between reconstructed ALP and the beam axis $\theta_{\gamma\gamma}$, the transverse momentum of reconstructed ALP $p_{T}^{\gamma\gamma}$, invariant mass of two photons $m_{\gamma\gamma}$ and the total transverse energy of the final states $E_T$ are taken to be analyzed  when the ALP mass is in the region of $25$ GeV to $130~(50)$ GeV, meanwhile the normalized distributions of them are given in Fig.~\ref{distribution:2} based on four benchmark points as $m_a = 30$, $60$, $90$, $120$ GeV at the $240$ GeV CEPC with $\mathcal{L}= 5.6$ ab$^{-1}$. The kinematic distributions at the $91$ GeV CEPC with $\mathcal{L}=$ $16$ ab$^{-1}$ are similar to those at the $240$ GeV CEPC with $\mathcal{L}=$ $5.6$ ab$^{-1}$, which are  not shown in this section and the following two sections.

According to the information of these kinematic distributions, improved cuts presented in Table~\ref{table:aevcut} are further imposed for separating the signal events from the background events, where the particle number in the final state is subject to the condition of $N_{\gamma} > 1$. In Table~\ref{tab:2} and Table~\ref{tab:3}, we summarized the cross sections of the signal and background after
taking the step-by-step cuts for few ALP mass benchmark points and the specific parameter $Tr(g_{a\nu\nu})/f_a= Tr(g_{a\ell\ell})/f_a=0.1$ GeV$^{-1}$ at the $240$ GeV CEPC with $\mathcal{L}=$ $5.6$ ab$^{-1}$. It can be seen that the background is effectively suppressed. We further calculate the statistical significance $SS=S/\sqrt{S + B}$ for the integrated luminosity of $5.6$ ab$^{-1}$, where $S$ and $B$ are the number of events for the signal and background, respectively. Large $SS$ values can be attained in a broad region of the parameter space, as illustrated in Table~\ref{tab:2} and Table~\ref{tab:3}. The statistical significance can reach $20.768$ ($7.034$) for $m_a=15$ GeV ($120$ GeV). The cross sections of the signal and background after taking the step-by-step cuts for few ALP mass benchmark points and the specific parameter at $91$ GeV CEPC with $\mathcal{L}=$ $16$ ab$^{-1}$ are not given specifically in this section and the following two sections.

\begin{table*}[!ht]
\centering
\scalebox{0.85}{
\begin{tabular}
[c]{c| c c c c c}\hline
\multirow{2}{*}{Cuts}    & \multicolumn{2}{c}{$\sqrt{s}=240$ GeV}  & \multicolumn{2}{c}{$\sqrt{s}=91$ GeV}  \\

	       &$10$ GeV $\leq$ $m_a\leq25$ GeV      &$25$ GeV $<$ $m_a \leq130$ GeV  ~~&$10$ GeV $\leq$ $m_a\leq25$ GeV &$25$ GeV $<$ $m_a \leq50$ GeV     \\ \hline
	Cut 1         &  $N_{\gamma}>1$                     &$N_{\gamma}>1$ ~~ &  $N_{\gamma}>1$                     &$N_{\gamma}>1$  \\
	Cut 2         & $p_{T}^{\gamma\gamma}>40$ GeV   & $0.6<\theta_{\gamma\gamma}<2.6$ ~~& $p_{T}^{\gamma\gamma}>20$ GeV   & $0.9<\theta_{\gamma\gamma}<2.2$  \\

	Cut 3         &  $\lvert m_{\gamma\gamma}- m_a\rvert\leq3$ GeV  &$p_{T}^{\gamma\gamma}>50$ GeV ~~&  $\lvert m_{\gamma\gamma}- m_a\rvert\leq3$ GeV  &$p_{T}^{\gamma\gamma}>22$ GeV   \\

	Cut 4         &  $\Delta R _{\gamma\gamma}<0.9$  & $\lvert m_{\gamma\gamma}- m_a\rvert\leq5$ GeV ~~&  $\Delta R _{\gamma\gamma}<2.3$  & $\lvert m_{\gamma\gamma}- m_a\rvert\leq5$ GeV   \\
    Cut 5         &  $-$               & $E_T>60$ GeV ~~&  $-$               & $E_T>30$ GeV    \\ \hline
\end{tabular}}
\caption{\label{table:aevcut}The improved cuts on signal and background for $10$ GeV $\leq$ $m_a\leq130~(50)$ GeV at the $240~(91)$ GeV CEPC.}
\end{table*}

\begin{table*}[!ht]
\centering
\setlength{\tabcolsep}{3mm}{
\resizebox{12cm}{!}{
\begin{tabular}{|c|cccc|}\hline
&\multicolumn{4}{c|}{cross sections for signal (background) [pb]} \\
\raisebox{0.8ex}{Cuts} & $m_a=10$ GeV & $m_a=15$ GeV &$m_a=20$ GeV& $m_a=25$ GeV \\ \hline
Basic Cuts & \makecell{$1.5939\times10^{-4}$\\$(0.06727)$} & \makecell{$6.2248\times10^{-4}$\\$(0.06727)$} &\makecell{$1.1256\times10^{-3}$\\$(0.06727)$} &\makecell{$1.1738\times10^{-3}$\\$(0.06727)$} \\ \hline
Cut 1 & \makecell{$6.1460\times10^{-5}$\\$(0.06555)$} & \makecell{$2.8932\times10^{-4}$\\$(0.06555)$} &\makecell{$6.7564\times10^{-4}$\\$(0.06555)$} &\makecell{$1.0429\times10^{-3}$\\$(0.06555)$} \\ \hline
Cut 2 & \makecell{$1.1880\times10^{-5}$\\$(0.03228)$} & \makecell{$2.3468\times10^{-4}$\\$(0.03228)$} &\makecell{$6.1486\times10^{-4}$\\$(0.03228)$} &\makecell{$9.6762\times10^{-4}$\\$(0.03228)$} \\ \hline
Cut 3 & \makecell{$1.1880\times10^{-5}$\\$(9.0550\times10^{-5})$} & \makecell{$2.3457\times10^{-4}$\\$(4.8572\times10^{-4})$} & \makecell{$6.1452\times10^{-4}$\\$(8.733\times10^{-4})$} &\makecell{$9.6657\times10^{-4}$\\$(1.1660\times10^{-3})$} \\ \hline
Cut 4 & \makecell{$1.1880\times10^{-5}$\\$(9.055\times10^{-5})$} & \makecell{$2.3457\times10^{-4}$\\$(4.7991\times10^{-4})$} &\makecell{$5.7957\times10^{-4}$\\$(6.7579\times10^{-4})$} &\makecell{$8.1423\times10^{-4}$\\$(5.6573\times10^{-4}$)} \\ \hline
$SS$  & $2.777$ & $20.768$ & $38.709$ & $51.869$ \\ \hline
\end{tabular}}}
\caption{\label{tab:2}After different cuts applied, the cross sections for the signal and background  at the 240 GeV CEPC with $\mathcal{L}=$ $5.6$ ab$^{-1}$ for $Tr(g_{a\nu\nu})/f_a=Tr(g_{a\ell\ell})/f_a=0.1$ GeV$^{-1}$ with the benchmark points $m_a = 10$, $15$, $20$, $25$ GeV.}
\end{table*}

\begin{table*}[!ht]
\centering
\setlength{\tabcolsep}{3mm}{
\resizebox{12cm}{!}{
\begin{tabular}{|c|cccc|}\hline
&\multicolumn{4}{c|}{cross sections for signal (background) [pb]} \\
\raisebox{0.8ex}{Cuts} & $m_a=30$ GeV & $m_a=60$ GeV &$m_a=90$ GeV& $m_a=120$ GeV \\ \hline
Basic Cuts & \makecell{$1.1731\times10^{-3}$\\$(0.06727)$} & \makecell{$1.1352\times10^{-3}$\\$(0.06727)$} &\makecell{$7.5978\times10^{-4}$\\$(0.06727)$} &\makecell{$3.0677\times10^{-4}$\\$(0.06727)$} \\ \hline
Cut 1 & \makecell{$1.0858\times10^{-3}$\\$(0.06555)$} & \makecell{$1.0874\times10^{-3}$\\$(0.06555)$} &\makecell{$7.0307\times10^{-4}$\\$(0.06555)$} &\makecell{$2.8498\times10^{-4}$\\$(0.06555)$} \\ \hline
Cut 2 & \makecell{$9.0795\times10^{-4}$\\$(0.03691)$} & \makecell{$8.5059\times10^{-4}$\\$(0.03691)$} &\makecell{$5.5323\times10^{-4}$\\$(0.03691)$} &\makecell{$2.2441\times10^{-4}$\\$(0.03691)$} \\ \hline
Cut 3 & \makecell{$9.0086\times10^{-4}$\\$(0.02256)$} & \makecell{$8.3038\times10^{-4}$\\$(0.02256)$} & \makecell{$4.8759 \times10^{-4}$\\$(0.02256)$} &\makecell{$1.1389\times10^{-4}$\\$(0.02256)$} \\ \hline
Cut 4 & \makecell{$9.0007\times10^{-4}$\\$(1.4728\times10^{-3})$} & \makecell{$8.2889\times10^{-4}$\\$(2.7381\times10^{-3})$} &\makecell{$4.8302\times10^{-4}$\\$(2.4496\times10^{-3})$} &\makecell{$9.405\times10^{-5}$\\$(1.0466\times10^{-3})$} \\ \hline
Cut 5 & \makecell{$8.7539\times10^{-4}$\\$(1.3277\times10^{-3})$} & \makecell{$8.1677\times10^{-4}$\\$(2.5977\times10^{-3})$} &\makecell{$4.7760\times10^{-4}$\\$(2.3279\times10^{-3})$} &\makecell{$9.112\times10^{-5}$\\$(8.4854\times10^{-4})$} \\ \hline
$SS$  & $44.134$ & $33.077$ & $21.337$ & $7.034$ \\ \hline
\end{tabular}}}
\caption{\label{tab:3}Same as Table~\ref{tab:2} but for the benchmark points $m_a = 30$, $60$, $90$, $120$ GeV.}
\end{table*}

\begin{figure}[h]
\centering 
\includegraphics[width=0.55\textwidth]{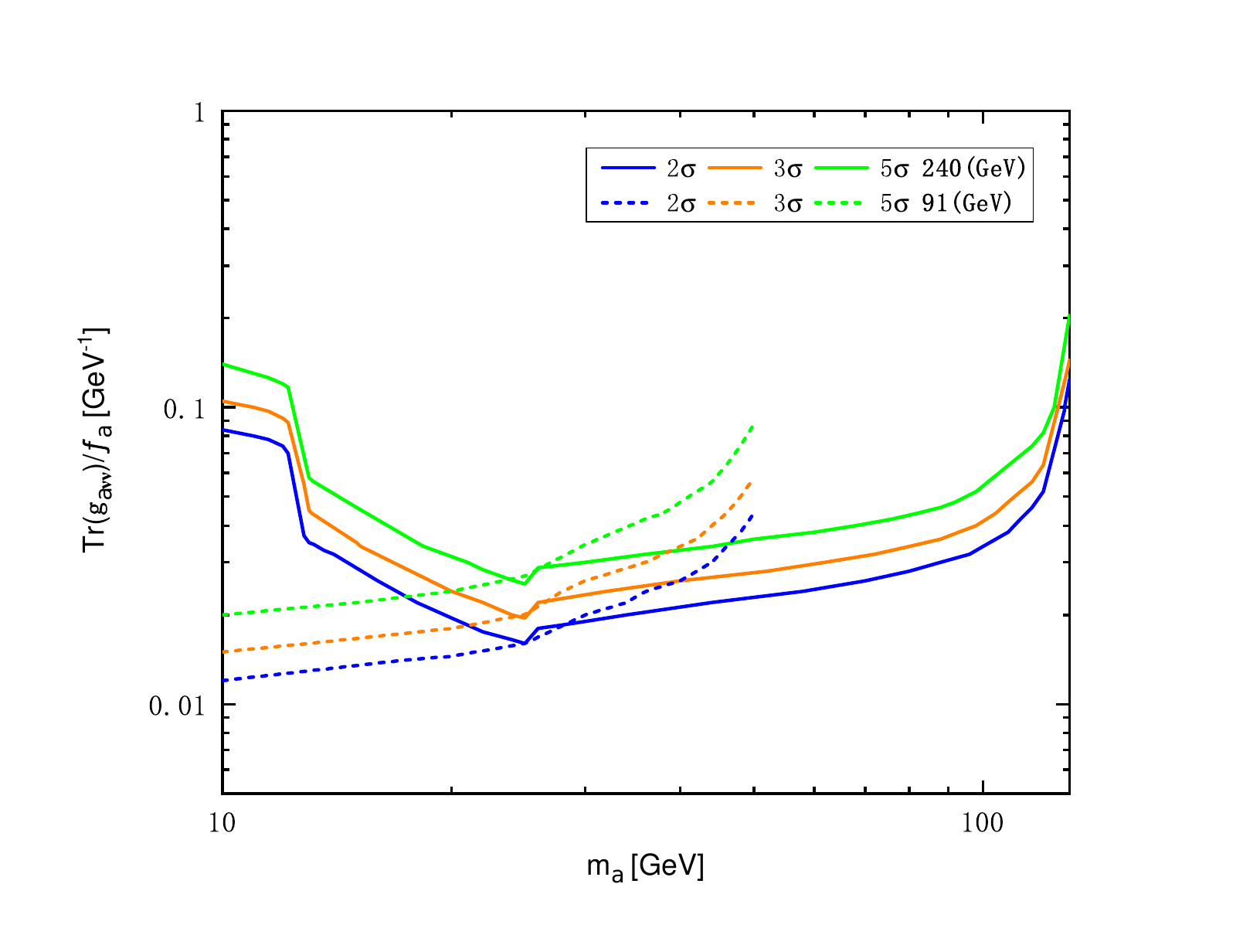}
\caption{For $c_{R}=0$, the $2\sigma$, $3\sigma$ and $5\sigma$ curves in the plane of $m_a - Tr(g_{a\nu\nu})/f_a$ from the signal process $e^{+}e^{-}\rightarrow\gamma\gamma\slashed E$ at the CEPC with $\sqrt{s}=240~(91)$ GeV and $\mathcal{L}=$ $5.6~(16)$ ab$^{-1}$.}
\label{fig:7}
\end{figure}

In Fig.~\ref{fig:7}, we plot the $2\sigma$, $3\sigma$ and $5\sigma$ curves in the plane of $m_a - Tr(g_{a\nu\nu})/f_a$ for the CEPC with $\sqrt{s}=240$~(91) GeV and $\mathcal{L}=$ $5.6$~(16) ab$^{-1}$. As shown in Fig.~\ref{fig:7}, we can obtain that the prospective sensitivities respectively as $0.016$~($0.012$) GeV$^{-1}$ $\leq$ $Tr(g_{a\nu\nu})/f_a$ $\leq$ $0.124$~($0.044$) GeV$^{-1}$, $0.0195$~($0.015$) GeV$^{-1}$ $\leq$ $Tr(g_{a\nu\nu})/f_a$ $\leq$ $0.144$~($0.06$) GeV$^{-1}$ and $0.0254$~($0.02$) GeV$^{-1}$ $\leq$ $Tr(g_{a\nu\nu})/f_a$ $\leq$ $0.204$~($0.088$) GeV$^{-1}$ at $2\sigma$, $3\sigma$ and $5\sigma$ levels in the ALP mass interval $10$ GeV $\sim$ $130$~($50$) GeV at the $240$~($91$) GeV CEPC with $\mathcal{L}=$ $5.6$~($16$) ab$^{-1}$.
The $91$ GeV CEPC can give a better expected sensitivity to the ALP-neutrino couplings when $m_a$ is smaller than about $25$ GeV, while the $240$ GeV CEPC can give a better prospective sensitivity when $m_a$ is larger than about $25$ GeV, due to the fact that the CEPC running at the $Z$ pole will produce a resonance enhancement, which results in a very large production cross section of the signal process, but in this case there is also a final state phase space depression. The resonance enhancement and the depression of the final state phase space partially cancel each other out, but within a limited mass range, making the observations at the $Z$ pole of the CEPC better than the $240$ GeV CEPC. In general, the CEPC might has the potential for detecting the ALP-neutrino couplings via their loop-level impact on the ALP couplings to EW gauge bosons in most of the mass range considered in this paper.

\begin{figure}[h]
\centering 
\includegraphics[width=0.48\textwidth]{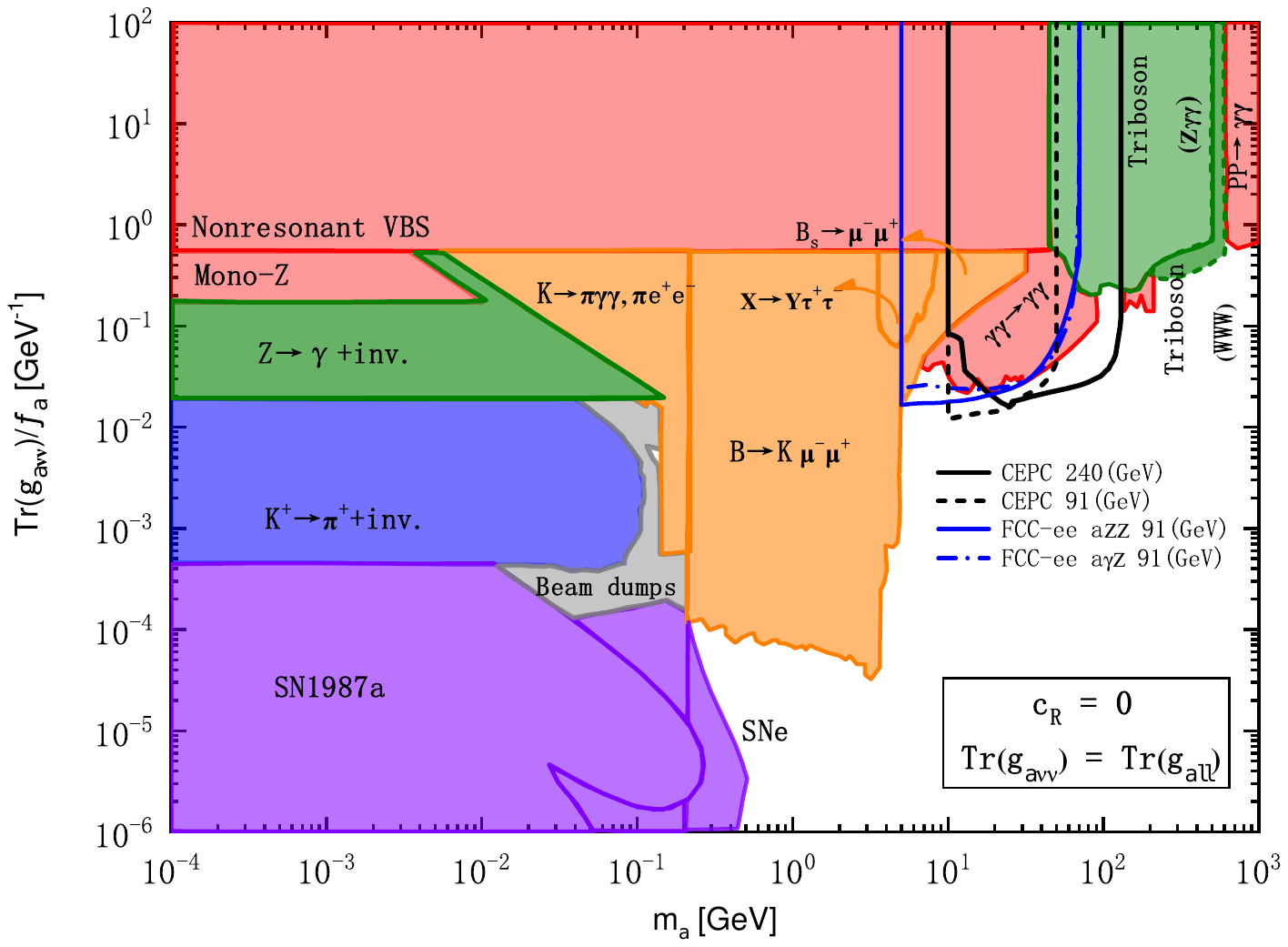}
\caption{For $c_{R}=0$, our projected $2\sigma$ sensitivities to the ALP-neutrino coupling $Tr(g_{a\nu\nu})/f_a$ at the CEPC with $\sqrt{s}=240~(91)$ GeV and $\mathcal{L}=$ $5.6~(16)$ ab$^{-1}$ in comparison with other current and expected excluded regions given in Ref.~\cite{Bonilla:2023dtf} and Ref.~\cite{Yue:2022ash}.}
\label{fig:8}
\end{figure}

In Fig.~\ref{fig:8}, we present our results for the projected sensitivity of the $240$~($91$) GeV CEPC with $\mathcal{L}=$ $5.6~(16)$ ab$^{-1}$ to the ALP-neutrino coupling at $95\%$ confidence level (C.L.)~(black solid and dashed lines) and other current and expected exclusion regions for this coupling. The current exclusion regions for the ALP-neutrino coupling come from the results of current collider experiments~(red and green regions), rare meson decays~(blue and orange regions), beam dump experiments~(grey region)~\cite{CHARM:1985anb,Riordan:1987aw,Blumlein:1990ay,Dolan:2017osp,NA64:2020qwq,Waites:2022tov} and astrophysical/cosmological observations~(purple region). The expected exclusion regions for the ALP-neutrino coupling come from the results of future collider experiments~(blue solid and dashed lines). The bounds on the ALP-neutrino coupling obtained by high-energy colliders mainly stem from the results of searches for ALPs via the process $Z\to \gamma +inv.$ at the LEP (labeled ``$Z\to\gamma+inv.$")~\cite{Craig:2018kne,L3:1997exg}, searches for mono-$Z$ events at the LHC (labeled ``Mono-$Z$")~\cite{Brivio:2017ije}, searches for non-resonant ALPs via Vector Boson Scattering~(VBS) processes at the LHC (labeled ``Nonresonant VBS")~\cite{Bonilla:2022pxu}, light-by-light ($\gamma\gamma \to \gamma\gamma$) scattering measured in Pb-Pb collisions at the LHC (labeled ``$\gamma\gamma \to \gamma\gamma$")~\cite{CMS:2018erd,ATLAS:2020hii}, diphoton production at the LHC (labeled ``$p p \to \gamma\gamma$")~\cite{Mariotti:2017vtv} and final triboson $WWW$ (labeled ``Triboson $(WWW)$")~\cite{CMS:2019mpq} and $Z\gamma\gamma$ (labeled ``Triboson $(Z\gamma\gamma)$")~\cite{Craig:2018kne} searches at the LHC, where the most stringent constraints on the ALP-neutrino coupling in the range $5$ GeV $\leq$ $m_a$ $\leq$ $90$ GeV come from the result of light-by-light ($\gamma\gamma \to \gamma\gamma$) scattering measured in Pb-Pb collisions at the LHC, for large ALP masses $m_a > m_Z$, the most stringent constraints on the ALP-neutrino coupling come from the result of diphoton production at the LHC. The blue and orange regions respectively depict the results from searches for ALPs via rare decays of the mesons $K$ and $B$. For light ALPs, the constraints on the ALP-neutrino coupling come from the decay process $K^+ \to \pi^+ \bar{\nu} \nu$ measured at the NA62 experiment (labeled ``$K^+ \to \pi^+ + inv.$")~\cite{NA62:2021zjw}. For intermediate ALP masses, many experimental measurements of meson decay processes can generate bounds on the ALP-neutrino coupling, such as the decay processes $K \to \pi a \to \pi\gamma\gamma ,\, \pi e^-e^+ $ (labeled ``$K \to \pi \gamma\gamma, \pi e^+e^- $")~\cite{E949:2005qiy,NA62:2014ybm,NA48:2002xke,KTeV:2008nqz}, $B \to K^{(*)} \mu^-\mu^+$ (labeled ``$B \to K \mu^-\mu^+$ ")~\cite{LHCb:2015nkv,LHCb:2016awg}, $B_s \to \mu^+ \mu^-$ (labeled ``$B_s \to \mu^- \mu^+$")~\cite{Albrecht:2019zul} and several decay processes mediated by $a \to\tau^-\tau^+$ (labeled ``$X \to Y \tau^+ \tau^-$")~\cite{BaBar:2012sau}, in which the most stringent limits on ALP-neutrino coupling arise from experimental measurements of the decay process $B \to K^{(*)} \mu^-\mu^+$. The purple regions depict the limits from experimental observations of supernovae~\cite{Ferreira:2022xlw,Diamond:2023scc,Caputo:2022mah}, which are labeled as ``SNe" and ``SN1987a". In addition, the blue solid and dashed lines respectively depict the expected sensitivities to the ALP-neutrino coupling from the processes $Z\to \mu^{+} \mu^{-} \slashed E$ and $Z\to e^+ e^- \mu^{+} \mu^{-}$ at the FCC-ee running at the $Z$ pole~\cite{Yue:2022ash}. It is important to note that some previous papers have investigated the expected sensitivity to the couplings of ALP with leptons, e.g., Ref.~\cite{Bauer:2018uxu} has obtained the expected sensitivity to the ALP coupling to leptons from their loop-level impact on the ALP couplings to electroweak gauge bosons via the processes $e^+ e^- \to \gamma a$~$(a\to \ell^+ \ell^-)$ and $e^+ e^- \to Z a$~$(a\to \ell^+ \ell^-)$ at the CLIC and the FCC-ee. By comparison, we find that the expected sensitivities obtained by Ref.~\cite{Bauer:2018uxu} are better than those given by the CEPC in this paper, but their work is based on the assumption of $Br(a\to \ell^+ \ell^-) = 1$. However,  ALP decays into two photons might be one of main decay channels when the ALP mass $m_a$ is small, and therefore this work is not shown on the plots in Fig.~\ref{fig:8}. Furthermore, Ref.~\cite{Calibbi:2022izs} has also investigated the sensitivity to leptonic ALP couplings at future $e^+$$e^-$ colliders and has given the expected exclusion regions for the ALP mass $m_a$ range up to $200$ MeV, which does not coincide with the ALP mass range considered in this paper, and as a consequence, this work is also not plotted in Fig.~\ref{fig:8}.

Comparing projected sensitivity derived in this work to the ALP-neutrino coupling $Tr(g_{a\nu\nu})/f_a$ with the regions excluded by other experiments, we can obtain a conclusion that the obtained $91$ GeV CEPC projected sensitivities are in the range of $0.012$ GeV$^{-1}$ $\sim$ $0.016$ GeV$^{-1}$ for $m_a$ in the range from $10$ GeV to $25$ GeV, which are not covered by the bounds given by light-by-light ($\gamma\gamma \to \gamma\gamma$) scattering measured in Pb-Pb collisions, searches for non-resonant ALPs in VBS processes at the LHC and the projected sensitivities given by the processes $Z\to \mu^{+} \mu^{-} \slashed E$ and $Z\to e^+ e^- \mu^{+} \mu^{-}$ at the $Z$ pole of the FCC-ee. The obtained $240 $ GeV CEPC projected sensitivities are in the range of $0.016$ GeV$^{-1}$ $\sim$ $0.124$ GeV$^{-1}$ for $m_a$ in the range from $25$ GeV to $130$ GeV, which are not covered by the bounds given by light-by-light ($\gamma\gamma \to \gamma\gamma$) scattering measured in Pb-Pb collisions, searches for non-resonant ALPs in VBS processes, final triboson $Z\gamma\gamma$ searches at the LHC and the projected sensitivities given by the processes $Z\to \mu^{+} \mu^{-} \slashed E$ and $Z\to e^+ e^- \mu^{+} \mu^{-}$ at the $Z$ pole of the FCC-ee. Thus, the CEPC with $\sqrt{s}=240$~$(91)$ GeV and $\mathcal{L}=$ $5.6$~$(16)$ ab$^{-1}$ has a better potential to detect the ALP-neutrino couplings or gives more severe limits in the ALP mass range from $25$~$(10)$ GeV to $130$~$(25)$ GeV.

\subsection{In the case $c_{L}=0$, $Tr(g_{a\nu\nu})=0$}\label{subsec2}

\begin{figure}[H]
\centering 
\includegraphics[width=0.55\textwidth]{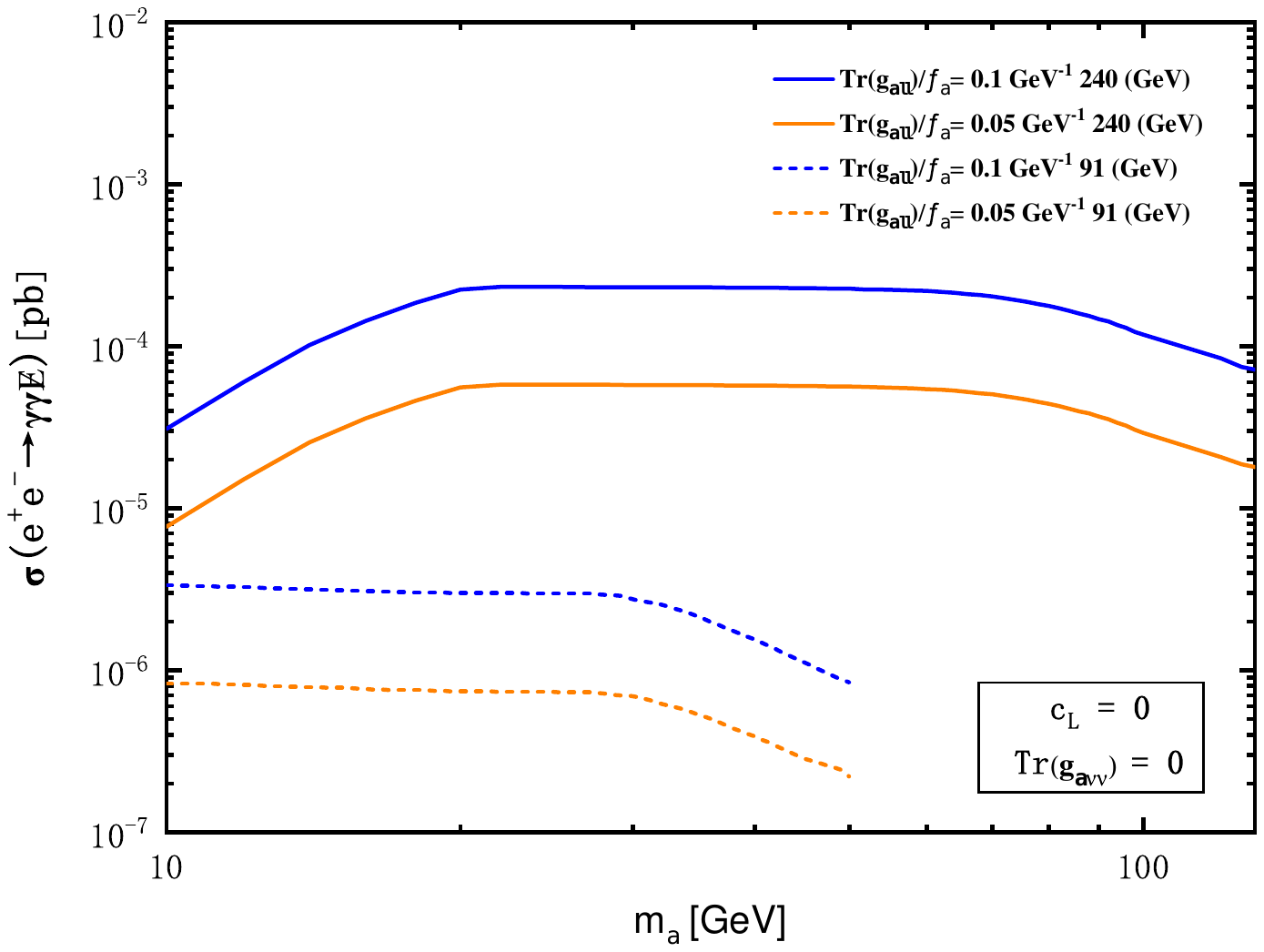}
\caption{The production cross section of the signal process $e^{+}e^{-}\rightarrow\gamma\gamma\slashed E$ as a function of the ALP mass $m_a$ at the $240$ GeV~(solid line) and the $91$ GeV~(dashed line) CEPC for $c_{L}=0$.}
\label{fig:9}
\end{figure}

In this subsection, we investigate the same signal process $e^{+}e^{-}\rightarrow\gamma\gamma\slashed E$ as mentioned above for $c_{L}=0$ with efforts to obtain projected sensitivity to the ALP-charged lepton coupling $Tr(g_{a\ell\ell})/f_a$. In this case, the first four graphs in Fig.~\ref{fig:2} have contributions to the signal process. Its production cross sections induced by ALPs with $10$ GeV $\leq m_a \leq$ $130$~$(50)$ GeV and different $Tr(g_{a\ell\ell})/f_a$ at the $240$~$(91)$ GeV CEPC are shown in Fig.~\ref{fig:9}, where the basic cuts have applied as previously stated. We can see that the image of the signal cross section has the same trend as that in Fig.~\ref{fig:4}, but the values of the cross section is smaller than that. This is due to that the contributions from the $W^{+}W^{-}$ fusion process are absent in this case. So, we still choose the $25$ GeV mass point as the breakpoint to divide the mass range of ALP into two intervals in this subsection. For $m_a \simeq$ $25$ GeV, the values of cross section are $5.8259\times10^{-5}$~$(7.589\times10^{-7})$ pb and $2.3388 \times 10^{-4}$~$(3.050\times10^{-6})$ pb for $Tr(g_{a\ell\ell})/f_a$ with $0.05$ GeV$^{-1}$ and $0.1$ GeV$^{-1}$ at the $240$~$(91)$ GeV CEPC, respectively, which are smaller than that of the corresponding SM background $0.06727$~($1.311\times10^{-4}$) pb. The kinematic distributions for the signal and background are similar to the case mentioned above. So we employ the same improved cut in this case. In Table~\ref{tab:4} and Table~\ref{tab:5}, we summarized the cross sections of the signal and background after imposing above improved cuts for few ALP mass benchmark points and the specific parameter $Tr(g_{a\nu\nu})/f_a=0, Tr(g_{a\ell\ell})/f_a=0.1$ GeV$^{-1}$ at the $240$ GeV CEPC with $\mathcal{L}=$ $5.6$ ab$^{-1}$. The corresponding $SS$ value of each benchmark point is also shown in the Table~\ref{tab:4} and Table~\ref{tab:5}. The statistical significance can reach $9.656$ ($1.519$) for $m_a=15$ GeV ($120$ GeV).

\begin{figure}[h]
\centering 
\includegraphics[width=0.55\textwidth]{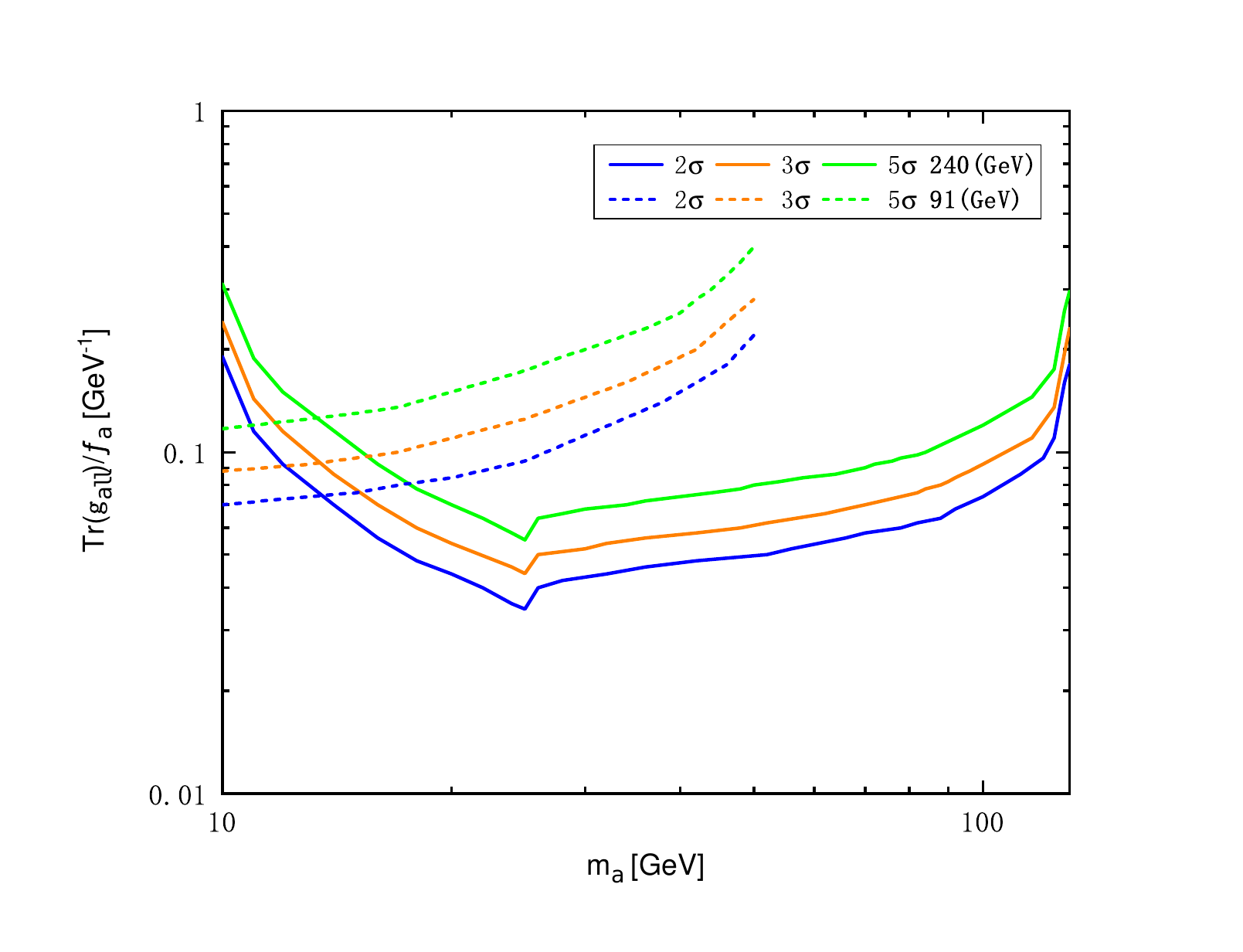}
\caption{For $c_{L}=0$, the $2\sigma$, $3\sigma$ and $5\sigma$ curves in the plane of $m_a - Tr(g_{a\ell\ell})/f_a$ from the process $e^{+}e^{-}\rightarrow\gamma\gamma\slashed E$ at the CEPC with $\sqrt{s}=240~(91)$ GeV and $\mathcal{L}=$ $5.6~(16)$ ab$^{-1}$.}
\label{fig:10}
\end{figure}

\begin{table*}[!ht]
\centering
\setlength{\tabcolsep}{3mm}{
\resizebox{12cm}{!}{
\begin{tabular}{|c|cccc|}\hline
&\multicolumn{4}{c|}{cross sections for signal (background) [pb]} \\
\raisebox{1.5ex}{Cuts} & $m_a=10$ GeV & $m_a=15$ GeV &$m_a=20$ GeV& $m_a=25$ GeV \\ \hline
Basic Cuts & \makecell{$3.1212\times10^{-5}$\\$(0.06727)$} & \makecell{$1.2271\times10^{-4}$\\$(0.06727)$} &\makecell{$2.2382\times10^{-3}$\\$(0.06727)$} &\makecell{$2.3291\times10^{-3}$\\$(0.06727)$} \\ \hline
Cut 1 & \makecell{$1.213\times10^{-5}$\\$(0.06555)$} & \makecell{$5.675\times10^{-5}$\\$(0.06555)$} &\makecell{$1.3346\times10^{-4}$\\$(0.06555)$} &\makecell{$2.0661\times10^{-4}$\\$(0.06555)$} \\ \hline
Cut 2 & \makecell{$2.36\times10^{-6}$\\$(0.03228)$} & \makecell{$4.611\times10^{-5}$\\$(0.03228)$} &\makecell{$1.2162\times10^{-4}$\\$(0.03228)$} &\makecell{$1.9189\times10^{-4}$\\$(0.03228)$} \\ \hline
Cut 3 & \makecell{$2.36\times10^{-6}$\\$(9.0550\times10^{-5})$} & \makecell{$4.607\times10^{-5}$\\$(4.8572\times10^{-4})$} & \makecell{$1.2154\times10^{-4}$\\$(8.733\times10^{-4})$} &\makecell{$1.9166\times10^{-4}$\\$(1.1660\times10^{-3})$} \\ \hline
Cut 4 & \makecell{$2.36\times10^{-6}$\\$(9.055\times10^{-5})$} & \makecell{$4.607\times10^{-5}$\\$(4.7991\times10^{-4})$} &\makecell{$1.1473\times10^{-4}$\\$(6.7579\times10^{-4})$} &\makecell{$1.6202\times10^{-4}$\\$(5.6573\times10^{-4}$)} \\ \hline
$SS$  & $0.578$ & $4.755$ & $9.656$ & $14.213$ \\ \hline
\end{tabular}}}
\caption{\label{tab:4}After different cuts applied, the cross sections for the signal and background  at the 240 GeV CEPC  with $\mathcal{L}=$ $5.6$ ab$^{-1}$ for $Tr(g_{a\nu\nu})/f_a=0, Tr(g_{a\ell\ell})/f_a=0.1$ GeV$^{-1}$ with the benchmark points $m_a = 10$, $15$, $20$, $25$ GeV.}
\end{table*}
\begin{table*}[!ht]
\centering
\setlength{\tabcolsep}{3mm}{
\resizebox{12cm}{!}{
\begin{tabular}{|c|cccc|}\hline
&\multicolumn{4}{c|}{cross sections for signal (background) [pb]} \\
\raisebox{1.5ex}{Cuts} & $m_a=30$ GeV & $m_a=60$ GeV &$m_a=90$ GeV& $m_a=120$ GeV \\ \hline
Basic Cuts & \makecell{$2.3256\times10^{-4}$\\$(0.06727)$} & \makecell{$2.1943\times10^{-4}$\\$(0.06727)$} &\makecell{$1.4809\times10^{-4}$\\$(0.06727)$} &\makecell{$8.2974\times10^{-5}$\\$(0.06727)$} \\ \hline
Cut 1 & \makecell{$2.1520\times10^{-4}$\\$(0.06555)$} & \makecell{$2.0327\times10^{-4}$\\$(0.06555)$} &\makecell{$1.37\times10^{-4}$\\$(0.06555)$} &\makecell{$7.757\times10^{-5}$\\$(0.06555)$} \\ \hline
Cut 2 & \makecell{$1.7959\times10^{-4}$\\$(0.03691)$} & \makecell{$1.5839\times10^{-4}$\\$(0.03691)$} &\makecell{$1.0746\times10^{-4}$\\$(0.03691)$} &\makecell{$6.101\times10^{-5}$\\$(0.03691)$} \\ \hline
Cut 3 & \makecell{$1.7884\times10^{-4}$\\$(0.02256)$} & \makecell{$1.5532\times10^{-4}$\\$(0.02256)$} & \makecell{$9.495 \times10^{-5}$\\$(0.02256)$} &\makecell{$3.398\times10^{-5}$\\$(0.02256)$} \\ \hline
Cut 4 & \makecell{$1.787\times10^{-4}$\\$(1.4728\times10^{-3})$} & \makecell{$1.55\times10^{-4}$\\$(2.7381\times10^{-3})$} &\makecell{$9.377\times10^{-5}$\\$(2.4496\times10^{-3})$} &\makecell{$1.948\times10^{-5}$\\$(1.0466\times10^{-3})$} \\ \hline
Cut 5 & \makecell{$1.7411\times10^{-4}$\\$(1.3277\times10^{-3})$} & \makecell{$1.528\times10^{-4}$\\$(2.5977\times10^{-3})$} &\makecell{$9.275\times10^{-5}$\\$(2.3279\times10^{-3})$} &\makecell{$1.891\times10^{-5}$\\$(8.4854\times10^{-4})$} \\ \hline
$SS$  & $10.63$ & $6.894$ & $4.461$ & $1.519$ \\ \hline
\end{tabular}}}
\caption{\label{tab:5}Same as Table~\ref{tab:4} but for the benchmark points $m_a = 30$, $60$, $90$, $120$ GeV.}
\end{table*}
In Fig.~\ref{fig:10}, we plot the $2\sigma$, $3\sigma$ and $5\sigma$ curves in the plane of $m_a - Tr(g_{a\ell\ell})/f_a$ for the CEPC with $\sqrt{s}=240$~(91) GeV and $\mathcal{L}=$ $5.6$~(16) ab$^{-1}$. As shown in Fig.~\ref{fig:10}, we can obtain that the prospective sensitivities respectively as $0.035$~($0.07$) GeV$^{-1}$ $\leq$ $Tr(g_{a\ell\ell})/f_a$ $\leq$ $0.18$~($0.22$) GeV$^{-1}$, $0.044$~($0.088$)  GeV$^{-1}$ $\leq$ $Tr(g_{a\ell\ell})/f_a$ $\leq$ $0.23$~($0.28$) GeV$^{-1}$ and $0.055$~($0.117$) GeV$^{-1}$ $\leq$ $Tr(g_{a\ell\ell})/f_a$ $\leq$ $0.295$~($0.4$) GeV$^{-1}$ at $2$$\sigma$, 3$\sigma$ and 5$\sigma$ levels in the ALP mass interval $10$ GeV $\sim$ $130$~($50$) GeV at the $240$~($91$) GeV CEPC with $\mathcal{L}=$ $5.6$~($16$) ab$^{-1}$.
As shown in Fig.~\ref{fig:10}, it is obvious that the $91$ GeV CEPC can give a better expected sensitivity to the ALP-charged lepton couplings when $m_a$ is smaller than about $12$ GeV, while the $240$ GeV CEPC can give a better prospective sensitivity when $m_a$ is larger than about $12$ GeV. So we can say that the CEPC has certain abilities to detect the ALP-charged lepton couplings in most of the mass range considered in this paper.

\begin{figure}[H]
\centering 
\includegraphics[width=0.48\textwidth]{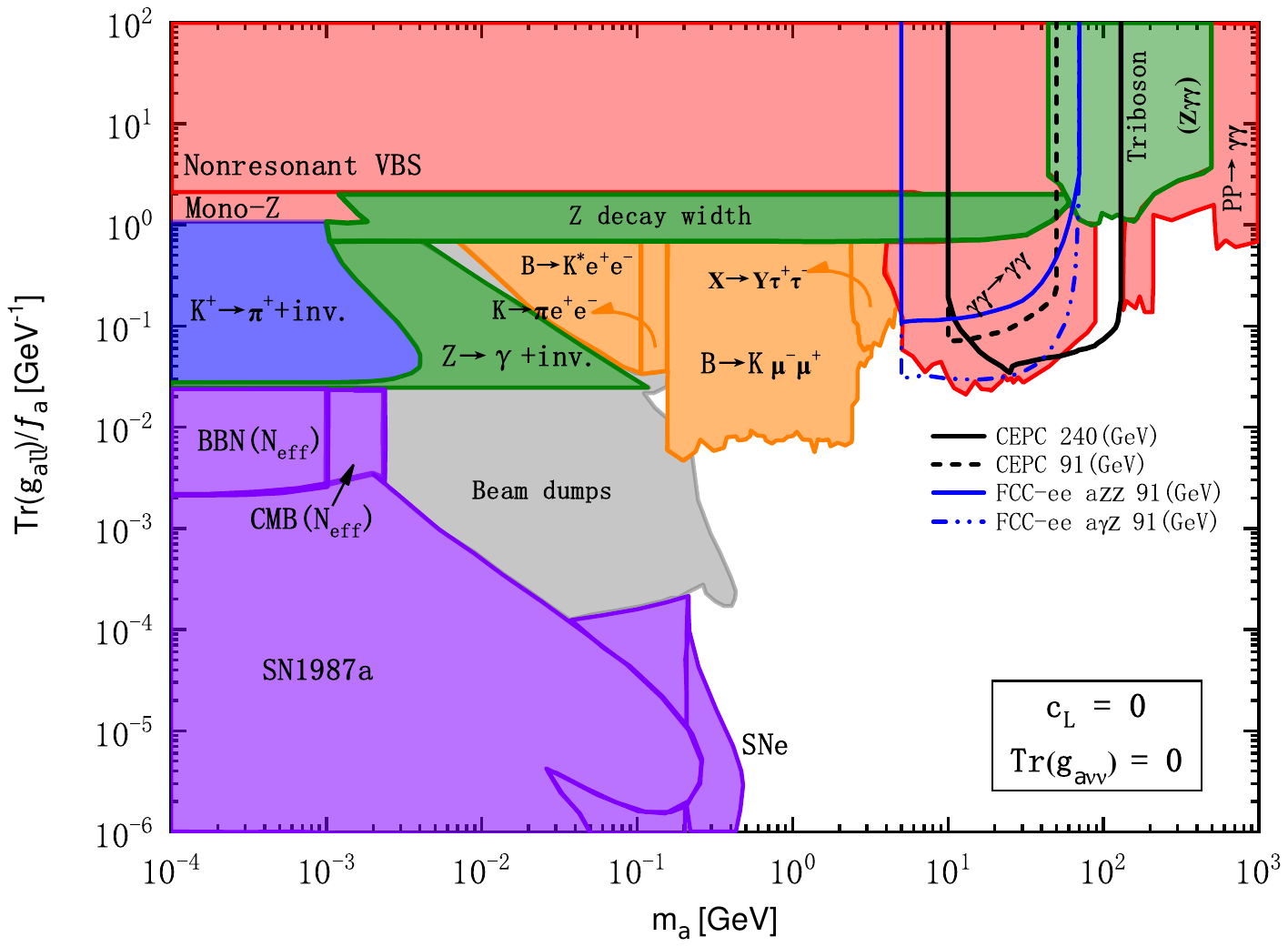}
\caption{For $c_{L}=0$, our projected $2\sigma$ sensitivities to the ALP-charged lepton coupling $Tr(g_{a\ell\ell})/f_a$ at the CEPC with $\sqrt{s}=240~(91)$ GeV and $\mathcal{L}=$ $5.6~(16)$ ab$^{-1}$ in comparison with other current and expected excluded regions given in Ref.~\cite{Bonilla:2023dtf} and Ref.~\cite{Yue:2022ash}.}
\label{fig:11}
\end{figure}

The projected sensitivities of the $240$~($91$) GeV CEPC with $\mathcal{L}=$ $5.6~(16)$ ab$^{-1}$to the ALP-charged lepton coupling at $95\%$ C.L.~(black solid and dashed lines) derived in this paper and current and expected exclusion regions for this coupling are given in Fig.~\ref{fig:11}. The green region depicts the limits from observations of the total $Z$ decay width at the LEP, which is labeled as ``$Z$ decay width"~\cite{Craig:2018kne,Brivio:2017ije}. The orange region gives the exclusion regions from the decay process $ B \to K^{*} e^{+} e^{-}$, which is labeled as ``$B \to K^{*} e^{+} e^{-}$"~\cite{LHCb:2015ycz}. In addition, the purple regions depict the bounds from CMB and BBN observations, which are labeled as ``CMB~($N_{eff}$)" and ``BBN~($N_{eff}$)"~\cite{Ghosh:2020vti,Depta:2020zbh}. Comparing prospective sensitivities derived in this work to the ALP-charged lepton coupling $Tr(g_{a\ell\ell})/f_a$ with the regions excluded by other experiments, we can come to a conclusion that the obtained $91$ GeV CEPC prospective sensitivities have been covered by the bounds given by light-by-light ($\gamma\gamma \to \gamma\gamma$) scattering measured in Pb-Pb collisions at the LHC and the prospective sensitivities given by the process $Z\to e^+ e^- \mu^{+} \mu^{-}$ at the $Z$ pole of the FCC-ee. While the obtained $240$ GeV CEPC prospective sensitivities are in the range of $0.052$ GeV$^{-1}$ $\sim$ $0.18$ GeV$^{-1}$ for $m_a$ in the range from $55$ GeV to $130$ GeV, which are not covered by the bounds given by light-by-light ($\gamma\gamma \to \gamma\gamma$) scattering measured in Pb-Pb collisions, searches for non-resonant ALPs in VBS processes, final triboson $Z\gamma\gamma$ searches at the LHC, observations of the total $Z$ decay width at the LEP and the prospective sensitivities given by the processes $Z\to \mu^{+} \mu^{-} \slashed E$ and $Z\to e^+ e^- \mu^{+} \mu^{-}$ at the $Z$ pole of the FCC-ee. Therefore, the CEPC with $\sqrt{s}=240$ GeV and $\mathcal{L}=$ $5.6$ ab$^{-1}$  has certain abilities to detect the ALP-charged lepton couplings in the ALP mass range from $55$ GeV to $130$ GeV in this paper.

\subsection{In the case $c_{L}=c_{R}$, $Tr(g_{a\ell\ell})=0$}\label{subsec3}

\begin{figure}[h]
\centering 
\includegraphics[width=0.52\textwidth]{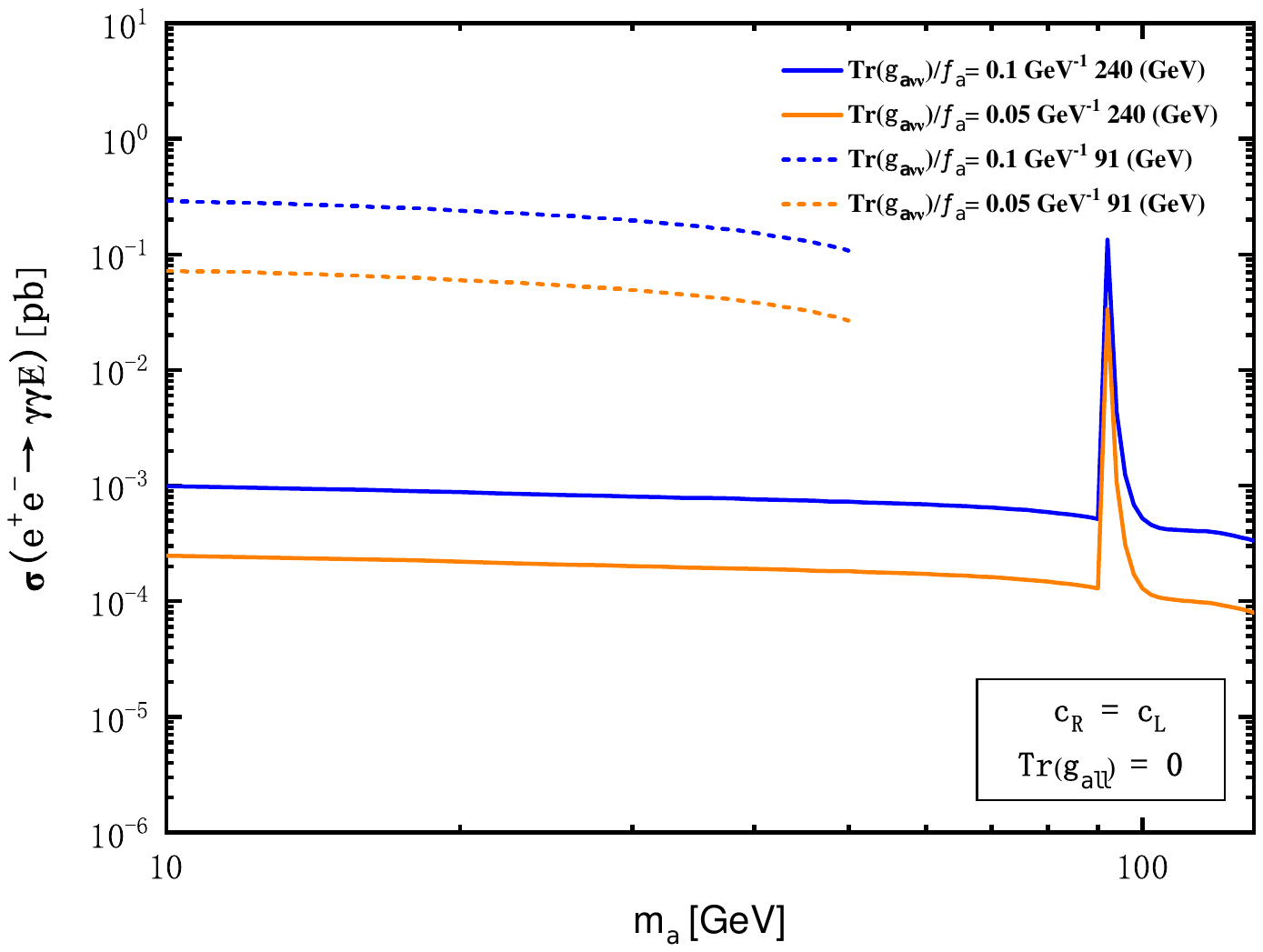}
\caption{The production cross section of the signal process $e^{+}e^{-}\rightarrow\gamma\gamma\slashed E$ as a function of the ALP mass $m_a$ at the $240$ GeV~(solid line) and the $91$ GeV~(dashed line) CEPC for $c_{L}=c_{R}$.}
\label{fig:12}
\end{figure}

In this subsection, we will consider the same signal process as described in the previous two subsections with the goal of obtaining prospective sensitivities to the ALP-neutrino coupling $Tr(g_{a\nu\nu})/f_a$ for $c_{L}=c_{R}$, in which case only the Fig.~\ref{fig:2}{b} has contributions to the signal process~\footnote{The ALP is off-shell when $m_a$ $<$ $m_Z$. In this case, there might be other potential loop contributions of the ALP to the signal, since the process cannot be factorised using the narrow width approximation. For example, an ALP is exchanged between the two initial¡ªstate electrons, and these processes appear in the perturbative expansion to a lower order than the product of two ALP-gauge boson couplings. However, all these processes will carry an sufficient fermion mass suppression to be reasonably ignored in this paper.}. Indeed, there may be other diagrams that contribute to the signal process, such as the process $e^+ e^- \to \nu_e \bar{\nu_e} \gamma Z \gamma$ ($Z \to \nu \bar{\nu}$), in which the ALP mediates the $Z\gamma$ final state and another photon is radiated from the interior of $W$ boson. However, the contributions of these diagrams to the signal are less than $0.01$$\%$, which is rationally ignored in this paper.
The production cross sections of the signal process $e^{+}e^{-}\rightarrow\gamma\gamma\slashed E$ induced by ALPs with $10$ GeV $\leq m_a \leq$ $130~(50)$ GeV and different $Tr(g_{a\nu\nu})/f_a$ at the $240$~(91) GeV CEPC are given in Fig.~\ref{fig:12}, where the basic cuts have applied as previously mentioned. We can see from Fig.~\ref{fig:12}, broadly speaking, the tendency of the cross section as varying with the ALP mass at the $240$ GeV CEPC is different from the first two cases mentioned above, in which the cross sections gradually decrease due to the final state phase space becomes limited with increasing ALP mass $m_a$, but there is a sharp peak when the mass of ALP approaches the mass of Z boson due to the resonance of $Z$ boson. At this point, the values of the cross section can reach $0.0335$ pb and $0.1339$ pb for $Tr(g_{a\nu\nu})/f_a$ equaling to $0.05$ GeV$^{-1}$ and $0.1$ GeV$^{-1}$, respectively. While the production cross section at the CEPC running at the $Z$ pole is larger than that at the $240$ GeV CEPC due to the considerable resonance enhancement in this case, which can reach $0.0719$ pb and $0.289$ pb for $Tr(g_{a\nu\nu})/f_a$ equaling to $0.05$ GeV$^{-1}$ and $0.1$ GeV$^{-1}$, respectively.

\begin{figure*}[!ht]
\centering
\subfigure[]{\includegraphics [scale=0.3]{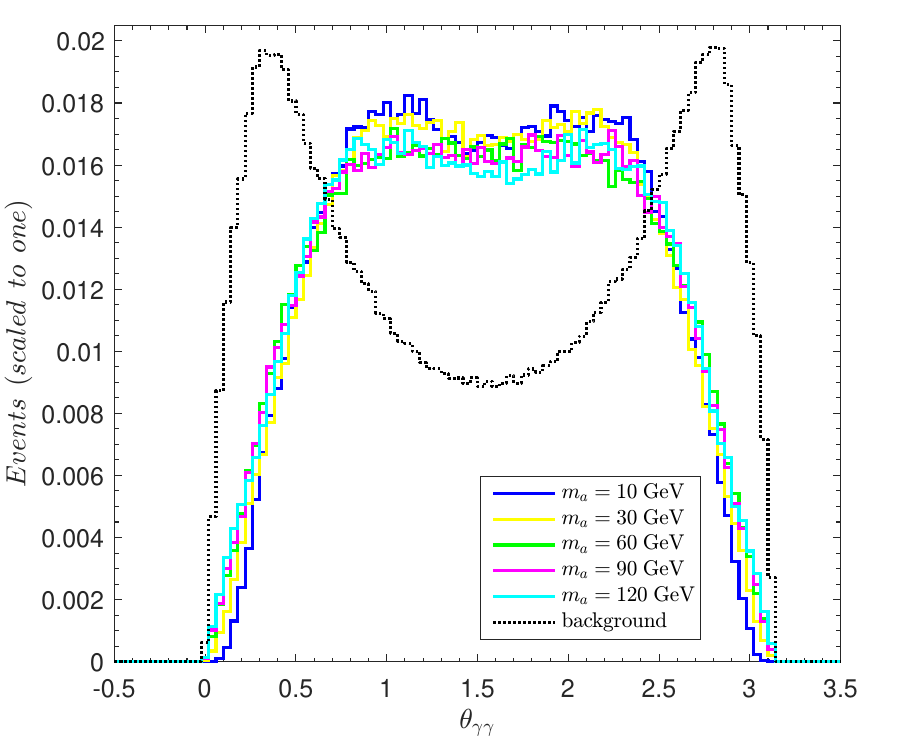}}
\hspace{0.2em}
\subfigure[]{\includegraphics [scale=0.3]{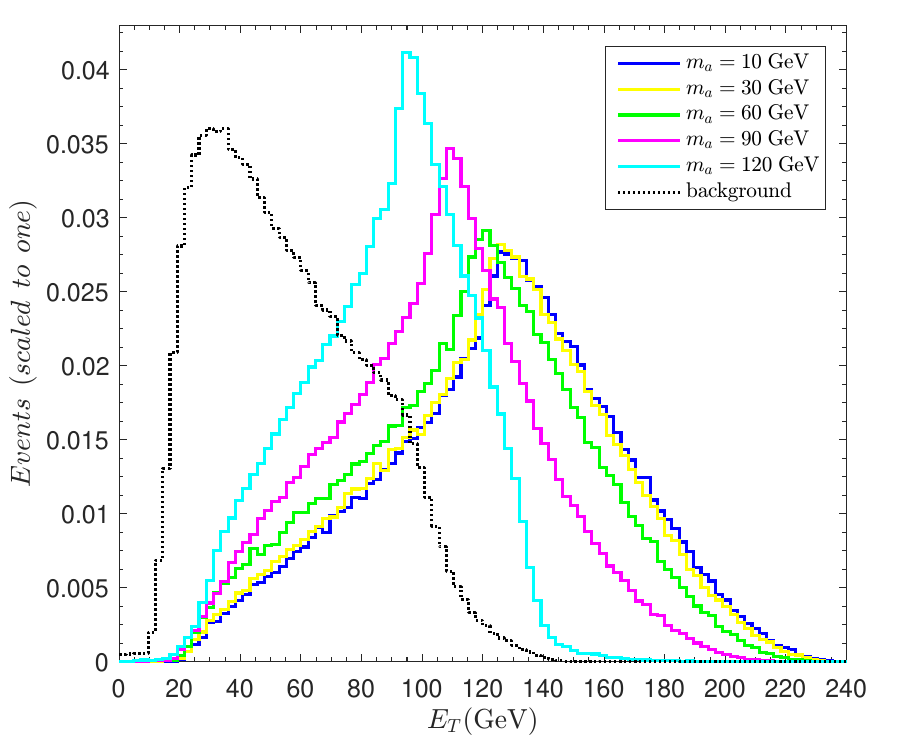}}
\hspace{0.2in}
\caption{The normalized distributions of the observables $\theta_{\gamma\gamma}$ (a) and $E_T$ (b) for the signal of selected ALP-mass benchmark points and SM background at the 240 GeV CEPC with $\mathcal{L}=$ $5.6$ ab$^{-1}$.}
\label{fig:13}
\end{figure*}

In the photophobic case, it can be seen that the two photons in the final state only come from the channel shown in Fig.~\ref{fig:2}{b}, therefore the kinematic variable of the invariant mass of the photon pair cannot better distinguish between signal and background, and we no longer choose this kinematic variable for our analyses. Variables of the angle between reconstructed ALP and the beam axis $\theta_{\gamma\gamma}$ and the total transverse energy of the final states $E_T$ are chosen for analysis. We present in Fig.~\ref{fig:13} the distribution of $\theta_{\gamma\gamma}$ and $E_T$ for the signal and background events with typical points of $m_a = 10$, $30$, $60$, $90$, $120$ GeV at the $240$ GeV CEPC with $\mathcal{L}=$ $5.6$ ab$^{-1}$.

Based on the kinematic distributions shown in Fig.~\ref{fig:13}, one set of optimized kinematical cuts listed in Table~\ref{table:6} are applied in the entire ALP mass interval of $10$ GeV to $130~(50)$ GeV to reduce backgrounds and improve the statistical significance. In Table~\ref{table:7}, we show the cross sections of the signal and background at the CEPC with $\sqrt{s}=240$ GeV and $\mathcal{L}=$ $5.6$ ab$^{-1}$ after imposing above cuts for few ALP mass benchmark points and the specific parameter $Tr(g_{a\ell\ell})/f_a=0, Tr(g_{a\nu\nu})/f_a = 0.1$ GeV$^{-1}$. The statistical significance can reach $20.791$ ($4.893$) for $m_a=10$ GeV ($120$ GeV).

\begin{table*}[!ht]
\centering
\setlength{\tabcolsep}{3mm}{
\begin{tabular}
[c]{c| c c}\hline
\multirow{2}{*}{Cuts}    & \multicolumn{1}{c}{$\sqrt{s}=240$ GeV}  & \multicolumn{1}{c}{$\sqrt{s}=91$ GeV}  \\

	       &~~~~~~~$10$ GeV $\leq$ $m_a\leq130$ GeV~~~~   &~~~~~~~$10$ GeV $\leq$ $m_a\leq50$ GeV~~~~     \\ \hline
	Cut 1         &  $N_{\gamma}>1$                 &  $N_{\gamma}>1$       \\
	Cut 2         & $0.6<\theta_{\gamma\gamma}<2.6$     & $0.9<\theta_{\gamma\gamma}<2.2$ \\

	Cut 3         &  $E_T>100$ GeV   &  $E_T>40$ GeV   \\ \hline

\end{tabular}}
\caption{\label{table:6}The improved cuts on signal and background for $10$ GeV $\leq$ $m_a\leq130~(50)$ GeV at the $240~(91)$ GeV CEPC.}
\end{table*}

\begin{table*}[!ht]
\centering
\resizebox{12cm}{!}{
\begin{tabular}{|c|ccccc|}\hline
&\multicolumn{5}{c|}{cross sections for signal (background) [pb]} \\
\raisebox{1.5ex}{Cuts}& $m_a=10$ GeV & $m_a=30$ GeV & $m_a=60$ GeV &$m_a=90$ GeV& $m_a=120$ GeV \\ \hline
Basic Cuts & \makecell{$9.8944\times10^{-4}$\\$(0.06727)$} & \makecell{$8.0754\times10^{-4}$\\$(0.06727)$} &\makecell{$6.8822\times10^{-4}$\\$(0.06727)$} &\makecell{$5.1417\times10^{-5}$\\$(0.06727)$} &\makecell{$3.9082\times10^{-5}$\\$(0.06727)$} \\ \hline
Cut 1 & \makecell{$9.5611\times10^{-4}$\\$(0.06555)$} & \makecell{$7.7452\times10^{-4}$\\$(0.06555)$} &\makecell{$6.5709\times10^{-4}$\\$(0.06555)$} &\makecell{$4.8923\times10^{-4}$\\$(0.06555)$} &\makecell{$3.7104\times10^{-4}$\\$(0.06555)$} \\ \hline
Cut 2 & \makecell{$8.025\times10^{-4}$\\$(0.03691)$} & \makecell{$6.4079\times10^{-4}$\\$(0.03691)$} &\makecell{$5.2562\times10^{-4}$\\$(0.03691)$} &\makecell{$3.9186\times10^{-4}$\\$(0.03691)$} &\makecell{$2.9621\times10^{-4}$\\$(0.03691)$}\\ \hline
Cut 3 & \makecell{$7.3336\times10^{-4}$\\$(6.2338\times10^{-3})$} & \makecell{$5.7334\times10^{-4}$\\$(6.2338\times10^{-3})$} & \makecell{$4.4204 \times10^{-4}$\\$(6.2338\times10^{-3})$} &\makecell{$2.8921\times10^{-4}$\\$(6.2338\times10^{-3})$} &\makecell{$1.6541\times10^{-4}$\\$(6.2338\times10^{-3})$}\\ \hline
$SS$  & $20.791$ & $16.445$ & $12.803$ & $8.474$ & $4.893$\\ \hline
\end{tabular}}
\caption{\label{table:7}After different cuts applied, the cross sections for the signal and background at the 240 GeV CEPC with $\mathcal{L}=$ $5.6$ ab$^{-1}$ for $Tr(g_{a\ell\ell})/f_a=0, Tr(g_{a\nu\nu})/f_a = 0.1$ GeV$^{-1}$ with the benchmark points $m_a = 10$, $30$, $60$, $90$, $120$ GeV.}
\end{table*}

\begin{figure}[h]
\centering 
\includegraphics[width=0.55\textwidth]{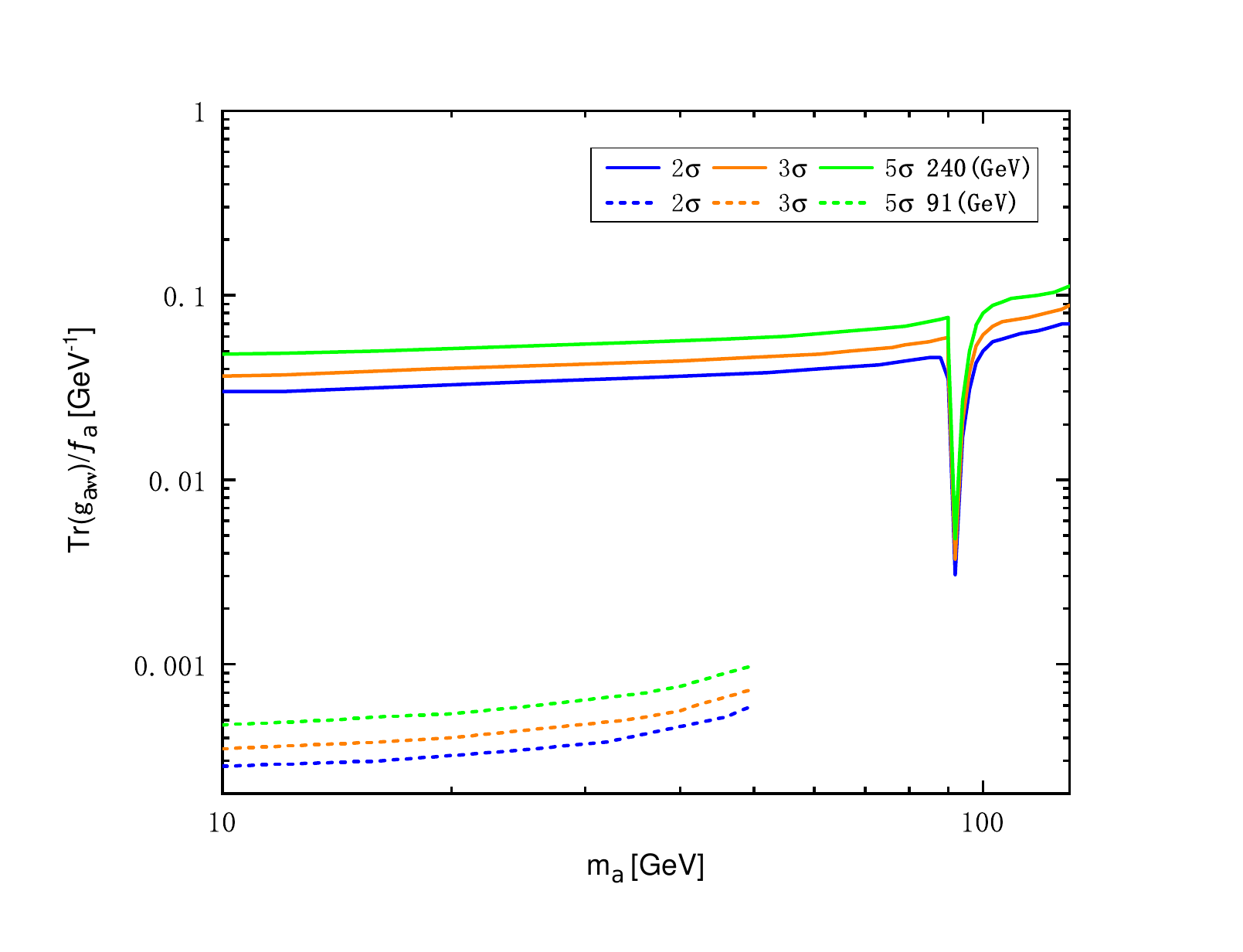}
\caption{For $c_{L}=c_{R}$, the $2\sigma$, $3\sigma$ and $5\sigma$ curves in the plane of $m_a - Tr(g_{a\nu\nu})/f_a$ from the process $e^{+}e^{-}\rightarrow\gamma\gamma\slashed E$ at the CEPC with $\sqrt{s}=240~(91)$ GeV and $\mathcal{L}=$ $5.6~(16)$ ab$^{-1}$.}
\label{fig:14}
\end{figure}

In Fig.~\ref{fig:14}, we plot the $2\sigma$, $3\sigma$ and $5\sigma$ curves in the plane of $m_a - Tr(g_{a\nu\nu})/f_a$ for the CEPC with $\sqrt{s}=240~(91)$ GeV and $\mathcal{L}=$ $5.6~(16)$ ab$^{-1}$. As shown in Fig.~\ref{fig:14}, we can obtain that the prospective sensitivities respectively as $0.0031$~$(2.8\times 10^{-4})$ GeV$^{-1}$ $\leq$ $Tr(g_{a\ell\ell})/f_a$ $\leq$ $0.07$~$(6\times 10^{-4})$ GeV$^{-1}$, $0.0037$~$(3.5\times 10^{-4})$ GeV$^{-1}$ $\leq$ $Tr(g_{a\ell\ell})/f_a$ $\leq$ $0.088$~$(7.4\times 10^{-4})$ GeV$^{-1}$ and $0.0048$~$(4.7\times 10^{-4})$ GeV$^{-1}$ $\leq$ $Tr(g_{a\ell\ell})/f_a$ $\leq$ $0.112$~$(9.9\times 10^{-4})$ GeV$^{-1}$ at $2$$\sigma$, $3$$\sigma$ and $5$$\sigma$ levels in the ALP mass interval $10$ GeV $\sim$ $130$~$(50)$ GeV. The CEPC running at the $Z$ pole is more sensitive than the CEPC running at $240$ GeV to detect the ALP-neutrino couplings in the ALP mass range from $10$ GeV to $50$ GeV, while the $240$ GeV CEPC has better prospects to explore the ALP-neutrino couplings in the ALP mass range from from $50$ GeV to $130$ GeV, due to the fact that the CEPC running at the $Z$ pole will produce a resonance enhancement, and in this case the depression of the final state phase space decreases, leading to a very large cross section of the signal process, making the observations at the $Z$ pole of the CEPC better than the $240$ GeV CEPC within a limited mass range. From these results, it is apparent that the CEPC has the prospect of detecting the ALP-neutrino couplings in certain mass intervals considered in this paper.

\begin{figure}[h]
\centering 
\includegraphics[width=0.48\textwidth]{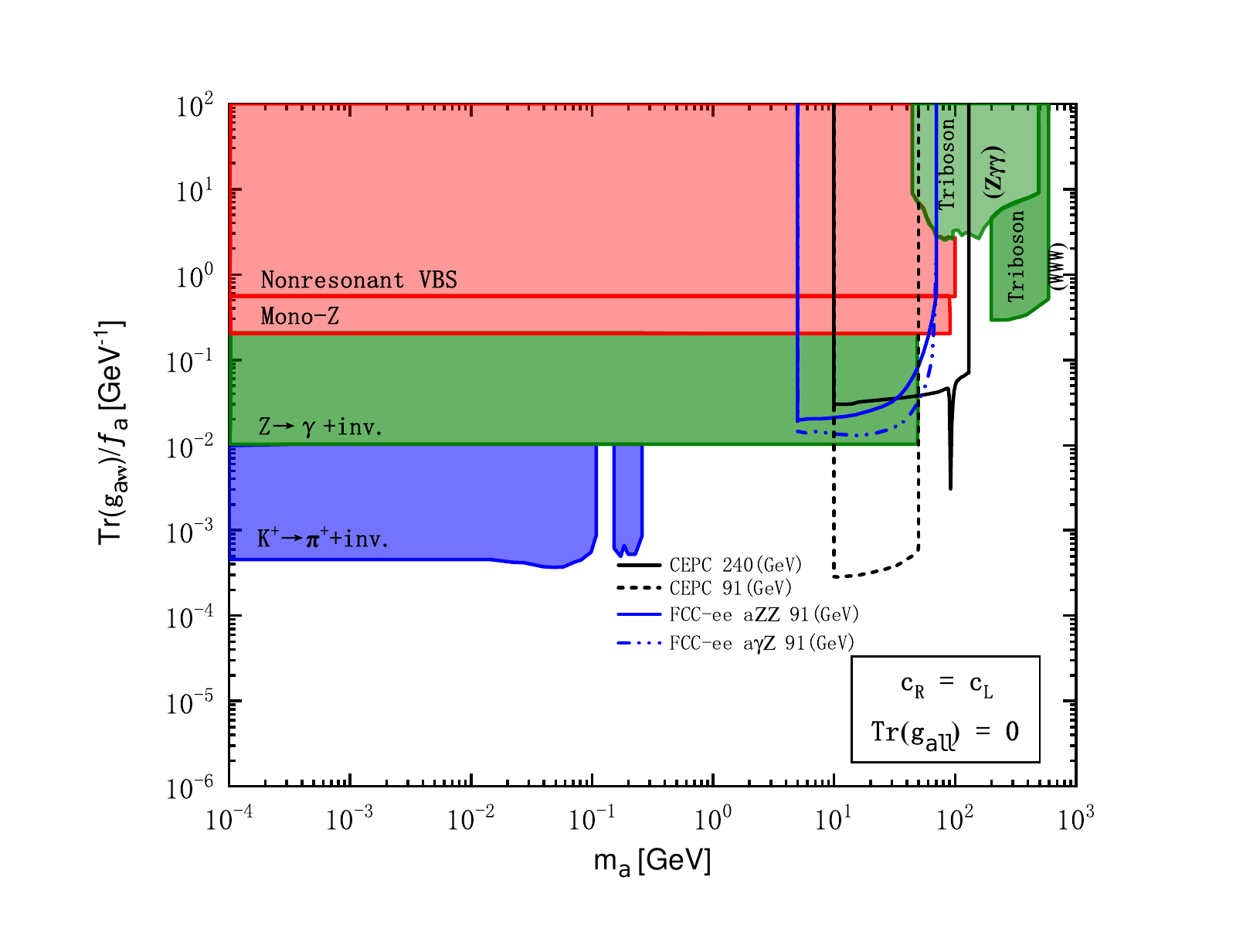}
\caption{For $c_{L}=c_{R}$, our projected $2\sigma$ prospective sensitivities to the ALP-neutrino coupling $Tr(g_{a\nu\nu})/f_a$ at the CEPC with $\sqrt{s}=240~(91)$ GeV and $\mathcal{L}=$ $5.6~(16)$ ab$^{-1}$ in comparison with other current and expected excluded regions given in Ref.~\cite{Bonilla:2023dtf} and Ref.~\cite{Yue:2022ash}.}
\label{fig:15}
\end{figure}

In Fig.~\ref{fig:15}, we present our results for the prospective sensitivities of the $240$~($91$) GeV CEPC with $\mathcal{L}=$ $5.6~(16)$ ab$^{-1}$ to the ALP-neutrino coupling at $95\%$ C.L.~(black solid and dashed lines) and other current and expected exclusion regions for this coupling in the photophobic case. Comparing prospective sensitivities derived in this work to the ALP-neutrino coupling $Tr(g_{a\nu\nu})/f_a$ with the regions excluded by other experiments, we find that the prospective sensitivities obtained at the $91$ GeV CEPC are in the range of $2.8\times 10^{-4}$ GeV$^{-1}$ $\sim$ $6\times 10^{-4}$ GeV$^{-1}$ in the ALP mass interval $10$ GeV $\sim$ $50$ GeV, which are not covered by the bounds given by searches for non-resonant ALPs in VBS processes, searches for mono-$Z$ events at the LHC, searches for ALPs via the process $Z\to \gamma +inv.$ at the LEP and the prospective sensitivities given by the processes $Z\to \mu^{+} \mu^{-} \slashed E$ and $Z\to e^+ e^- \mu^{+} \mu^{-}$ at the $Z$ pole of the FCC-ee. Besides, we find that the prospective sensitivities obtained at the $240$ GeV CEPC are in the range of $0.0031$ GeV$^{-1}$ $\sim$ $0.07$ GeV$^{-1}$ in the ALP mass interval $50$ GeV $\sim$ $130$ GeV, which are not covered by the bounds given by searches for non-resonant ALPs in VBS processes, searches for mono-$Z$ events, final triboson $Z\gamma\gamma$ searches at the LHC and the prospective sensitivities given by the processes $Z\to \mu^{+} \mu^{-} \slashed E$ and $Z\to e^+ e^- \mu^{+} \mu^{-}$ at the $Z$ pole of the FCC-ee. It is obvious that the CEPC with $\sqrt{s}=240$~$(91)$ GeV and $\mathcal{L}=$ $5.6$~$(16)$ ab$^{-1}$ has the prospect of detecting the couplings of ALP with neutrinos in the ALP mass range from $50$~$(10)$ GeV to $130$~$(50)$ GeV.

\section{Summary and discussion}\label{sec4}
The axion or axion-like particle is one of the best motivated particle candidates beyond the SM. Its rich phenomenology has been extensively studied in various low- and high energy  experiments. However, there are relatively few studies on the ALP-neutrino couplings at collider experiments due to the small neutrino mass. Owing to the loop-level impact of the ALP-lepton couplings on the ALP-EW gauge boson couplings, the current experimental limits or the prospective constraints of future experiments on the ALP couplings with EW gauge bosons can be converted to limits or prospective limits on the effective ALP-lepton couplings. Therefore, in this work, we consider some physical processes with cleaner backgrounds induced by the ALP couplings to EW gauge bosons for study.

In this paper, we have investigated the possibility of detecting the couplings of ALP with leptons from their loop-level impact on the ALP couplings to EW gauge bosons via the signal process $e^{+}e^{-}\rightarrow\gamma\gamma\slashed E$ at the $240$~(91) GeV CEPC with $\mathcal{L}=$ $5.6~(16)$ ab$^{-1}$ in three cases. We have performed the numerical calculations and the phenomenological analysis for the signal and relevant SM background and obtained the $2$$\sigma$, $3$$\sigma$ and $5$$\sigma$ prospective sensitivities to the ALP coupling parameters $Tr(g_{a\nu\nu})/f_a$ and $Tr(g_{a\ell\ell})/f_a$ at the CEPC with $\sqrt{s}=240~(91)$ GeV and $\mathcal{L}=$ $5.6~(16)$ ab$^{-1}$ in three cases. Our numerical results show that the CEPC running at the $Z$ pole is more sensitive than the CEPC running at $240$ GeV to detect the ALP-neutrino coupling $Tr(g_{a\nu\nu})/f_a$ in the ALP mass range from $10$ GeV to $25~(50)$ GeV for $c_{R}=0$~($c_{L}=c_{R}$) and the ALP-charged lepton coupling $Tr(g_{a\ell\ell})/f_a$ in the ALP
mass range from $10$ GeV to $12$ GeV for $c_{L}=0$, while the observations at the $240$ GeV CEPC are better than those at the $Z$ pole of the CEPC in the ALP mass range from $25~(50)$ GeV to $130$ GeV for $c_{R}=0$~($c_{L}=c_{R}$) and from $12$ GeV to $130$ GeV for $c_{L}=0$. Comparing our numerical results with other existing bounds and prospective sensitivities, the results we obtained show that the $240$~(91) GeV CEPC can cover a part of the ALP parameter space that are not excluded by current and future experiments. Specifically, the prospective sensitivities to the ALP-neutrino coupling $Tr(g_{a\nu\nu})/f_a$ are in the range of $0.016~(0.012)$ GeV$^{-1}$ $\sim$ $0.124~(0.016)$ GeV$^{-1}$ for $c_{R}=0$ and $25~(10)$ GeV $\leq$ $m_a$ $\leq$ $130~(25)$ GeV, $0.0031~(2.8 \times 10^{-4})$ GeV$^{-1}$ $\sim$ $0.07~(6 \times 10^{-4})$ GeV$^{-1}$ for $c_{L}=c_{R}$ and $50~(10)$ GeV $\leq$ $m_a$ $\leq$ $130~(50)$ GeV and to the ALP-charged lepton coupling $Tr(g_{a\ell\ell})/f_a$ are in the range of $0.052$ GeV$^{-1}$ $\sim$ $0.18$ GeV$^{-1}$ for $c_{L}=0$ and $55$ GeV $\leq$ $m_a$ $\leq$ $130$ GeV at the $240$~(91) GeV CEPC with $\mathcal{L}=$ $5.6~(16)$ ab$^{-1}$.

ALPs are usually studied within the framework of effective field theory (EFT). However, it is important to know whether the ALP Lagrangian is a valid and self-consistent EFT. The ALP EFT is required to satisfy the unitarity constraints, thus there have been many studies on the validity of the ALP EFT. For example, Ref.~\cite{Brivio:2021fog} has derived constraints on the effective interactions of ALP with the SM particles from partial-wave unitarity in 2 $\to$ 2 scattering. The prospective sensitivities to the ALP-lepton couplings given in this paper satisfy the unitarity constraints given by Ref.~\cite{Brivio:2021fog}.

In this paper, we choose the promising signal process $e^+ e^- \to \gamma \gamma \slashed E$ to probe the ALP-lepton couplings due to the simple signal and the clean background, which facilitates our signal and background analysis. Certainly, the couplings of ALP with leptons can also be explored via the process $e^+ e^- \to \ell^+ \ell^- \slashed E$, in which the final state $\ell^+$$\ell^-$ originates from the ALP, $\ell = e, \mu,\tau$. However, this process involves a large background and a complex signal process. Especially, due to the rapid decay of $\tau$, making signal identification very challenging, investigating the couplings of ALP with leptons through the process $e^+ e^- \to \ell^+\ell^- \slashed E$ needs to be further investigated, which will be one of our future works.

Our estimates of the prospective sensitivities at the CEPC are complementary to the existing bounds and the expected bounds given by the LHC, LEP and FCC-ee. Therefore the CEPC has a better potential to detect the ALP-lepton couplings or gives more severe limits. Hopefully this work will provide valuable insights for future collider experiments aiming to explore the ALP-lepton couplings via some processes with the simple signal and the clean background and be useful for explaining dark matter, astrophysical and cosmological phenomena in the future.

\section*{ACKNOWLEDGMENT}

This work was partially supported by the National Natural Science Foundation of China under Grant No. 11875157 and No. 12147214.

\bibliographystyle{sn-mathphys-num}
\bibliography{ALPref}

\end{document}